\documentclass[a4paper,11pt]{article}
\pdfoutput=1 

\usepackage{jcappub} 

\usepackage[T1]{fontenc} 
\usepackage{longtable}
\usepackage{array} 
\usepackage{multirow}

\title{Alborz-I array: a simulation on performance and properties of the array around the knee of the cosmic ray spectrum}

\author [a] {S. Abdollahi,}
\author [a,b,1] {M. Bahmanabadi,\note{Corresponding author.}}
\author [a,b] {Y. Pezeshkian,}
\author [a] {and S. Mortazavi Moghaddam}

\affiliation[a] {Alborz Observatory, Sharif University of Technology,\\ PO Box 11155-9161, Tehran, Iran}
\affiliation[b] {Department of Physics, Sharif University of Technology,\\ PO Box 11155-9161, Tehran, Iran}

\emailAdd{abdollahi.soheila@gmail.com}
\emailAdd{bahmanabadi@sharif.edu}
\emailAdd{pezeshkian@sut.ac.ir}
\emailAdd{mortazavimoghaddam@gmail.com}

\abstract{The first phase of the Alborz Observatory Array (Alborz-I) consists of 20 plastic scintillation detectors each one with surface area of 0.25 $m^{2}$
spread over an area of 40$\times$40 $m^{2}$ realized to the study of Extensive Air Showers around the $\it knee$ at the Sharif University of Technology campus. The first
stage of the project including construction and operation of a prototype system has now been completed and the electronics that will be used in the array
instrument has been tested under field conditions. In order to achieve a realistic estimate of the array performance, a large number of simulated CORSIKA~\cite{a} showers have been used. In the present work, theoretical results obtained in the study of different array layouts and trigger conditions are described. 
Using Monte Carlo simulations of showers
the rate of detected events per day and the trigger probability functions, i.e., the probability for an extensive air
shower to trigger a ground based array as a function of the shower core distance to the center of array are presented for energies above 1 TeV and zenith angles up to 60$^{\circ}$. Moreover, the angular resolution of the Alborz-I array is obtained.}

\keywords{ground based detector array, extensive air shower, trigger probability, angular resolution}
\begin{document}
\maketitle
\flushbottom

\section{Introduction}
\label{sec:1}
Study of the primary cosmic rays provides important information on their origins and acceleration mechanisms. The energy spectrum of cosmic rays follows a
power law and shows a bend around 3$\times$10$^{15}$ eV, known as the $\it knee$ region, where the spectral index changes from -2.7 to approximately -3.1~\cite{b}.
At these energies, indirect measurements of cosmic rays due to their low flux are performed using ground based arrays of particle detectors.
In the last decade, findings of many high-profile experiments confirm the general view which attributes the $\it knee$ to processes of magnetic confinement occurring
either at acceleration regions or as diffusive leakage from the Galaxy or both. \\

Air shower measurements can determine the arrival direction, energy, and mass of the primary cosmic ray.
A particle detector array provides statistical samples of secondary particles in the shower disc, and the resolution achieved in the shower reconstruction depends
on the sampling fraction of secondary particles from the shower disc~\cite{c}. \\

Lowest and highest energies at which cosmic rays can be detected by arrays depend on the distance between neighboring detectors and overall size of detector
array respectively. Therefore geometry and design of an array have an important role to extend the detectable energy range. In this paper, geometrical effects of
array layout on the number of detected events, trigger probability and angular resolution are studied by increasing distance between neighboring detectors and
also changing configuration of the array. \\

In order to show dependence of the trigger probability on the primary particle energy ($\it E$) and shower core distance from the array center ($\it r$), detailed studies on the trigger
rate for two rectangular and cluster layouts are carried out. The trigger probability of the simulated array as a function of $\it r$ and $\it E$, known as
the trigger probability function $\it P(r,E)$, is presented as a fundamental parameter to compare different trigger conditions in the both rectangular and cluster
layouts. On the other hand, the angular resolution of the array for different trigger conditions and layouts is calculated using a chi-squared minimization
algorithm. Two different shower sets simulated by CORSIKA, which differ in the distribution of energy and zenith angles, have been applied for these studies.

\section{The Alborz-I experiment}
The Alborz Observatory Array (AOA) has been designed to measure the flux, arrival direction and energy of the primary cosmic ray around the $\it knee$.
The AOA has been planned to have both Scintillation Detectors (SDs) and Water Cherenkov Detectors (WCDs). \\

The Alborz-I as the first phase of the AOA is realized at Sharif University of Technology (SUT), Tehran (35$^{\circ}$43$^{\prime}$N 51$^{\circ}$20$^{\prime}$E, 1200 m a.s.l corresponding to an average atmospheric depth of 890 gcm$^{-2}$), and consists of 20 scintillation detectors (each one 0.25 m$^{2}$) covering an
area of about 1600 m$^{2}$. In the next phase, 10 WCDs will be added to it.
The used detectors are optimized by a series of experiments and simulations~\cite{d,e}. \\

Each scintillation detector of the Alborz-I has a pyramidal galvanized iron light enclosure with 1 mm wall thickness and diffuse white coated inner surface,
housing a 50$\times$50$\times$2 cm$^{3}$ NE102A scintillator. Each detector is embedded in a station with a 5 mm cement ceiling. Scintillation light emitted due to passing a
particle through the scintillator is read out from tip of the pyramidal light enclosure by a 5 cm photomultiplier tube (PMT, 9813B) at the distance of 20 cm of the
scintillator sheet for timing measurements.\\

\begin{figure}[tbp]
\centering 
\includegraphics[width=.45\textwidth]{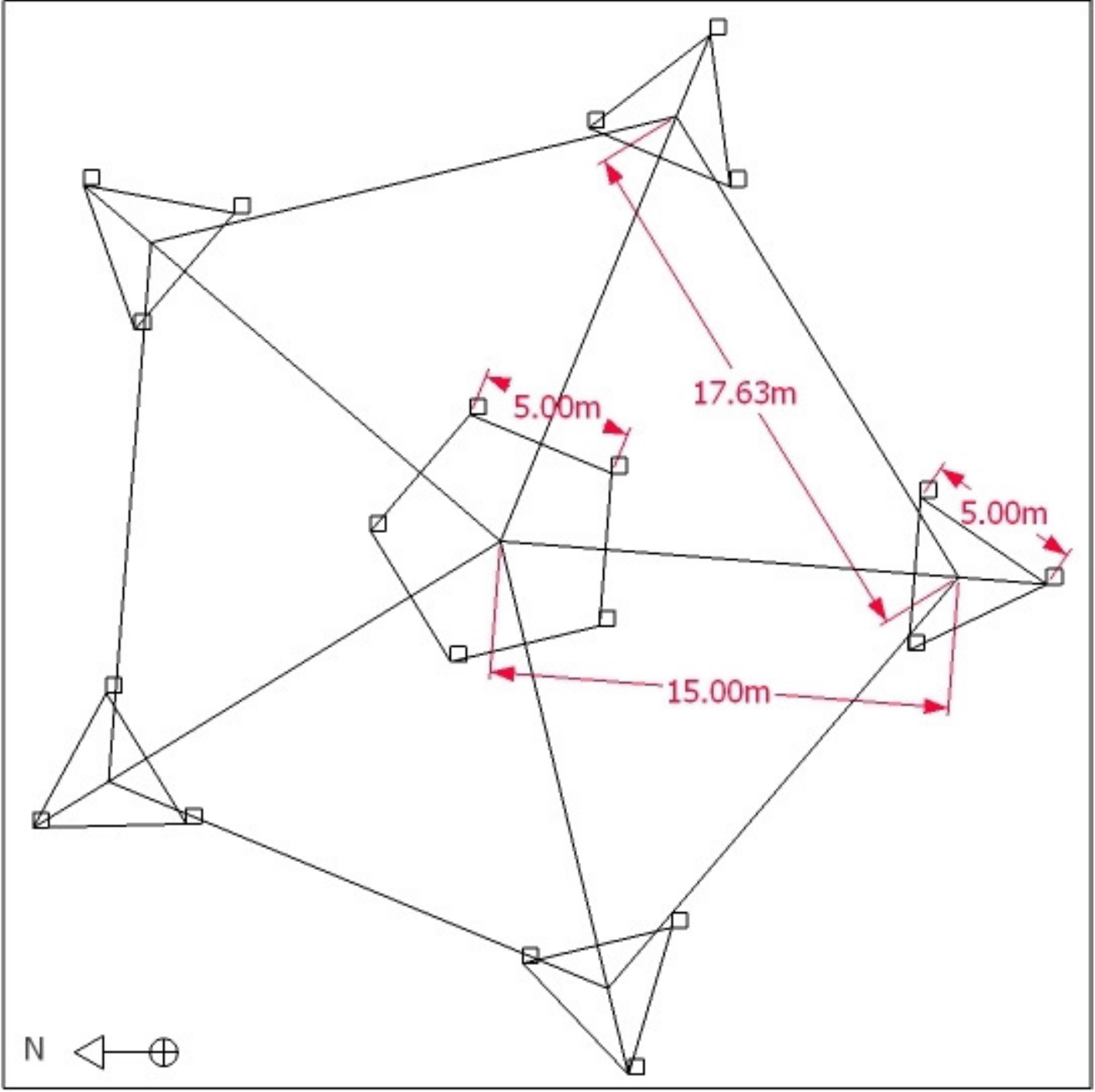}
\caption{\label{fig:1} Cluster layout of the Alborz-I array where 20 SDs will be deployed in a square area. Distance between neighboring detectors in the central pentagon and also in the surrounding triangles around the central pentagon is 5 m.}
\end{figure}

Two different array configurations, one cluster and other rectangular grid, are proposed. As shown in figure~\ref{fig:1}, in the cluster layout, 5 out of 20 detectors
arranged on an internal pentagon with side length 5 m and the rest set on 5 surrounding triangles with the same side length 5 m which are placed on the corners 
of an external pentagon with side length about 18 m.\\

In selection chain of the array properties, such as array layout, trigger condition and energy range of interest, several parameters including the trigger probability function, the
number of recorded events and the angular resolution of the array are studied for different implemented conditions in the simulations.
The results from study of the array properties are mentioned in the following sections.

\section{Monte Carlo simulations}
A realistic estimate of the array response and angular resolution, as two main characteristics of an array, requires a detailed knowledge of the physics of shower
development and interaction of secondary particles with the detector material. \\

Two different EAS sets simulated by CORSIKA (version 6.9), which differ in the distribution of energy and zenith angles, have been used.
QGSJET-II~\cite{f} and GHEISHA ($\it E\leqslant$80 GeV)~\cite{g} models have been employed for high and low energy hadronic interactions, respectively.
At high energies, the showers have been generated in the High Performance Computing Center (HPCC) of SUT with respect to time consuming simulations. \\

In order to reliable estimate of the trigger probability function, a shower set consists of 3600 extensive air showers with a composition of 88$\%$ proton and 12$\%$ alpha as primary particles have been simulated.  In this set, air showers were simulated with azimuth angles from 0$^{\circ}$ to 360$^{\circ}$ and zenith angles between 0$^{\circ}$ and 60$^{\circ}$ distributed as $\it I$$\propto$ $\sin$$\theta$ $\cos$$\theta$ d$\theta$ (which the $\it sine$ term respects the solid angle element of the sky, while the $\it cosine$ term takes into account the geometrical efficiency of a flat horizontal detector).  The energy of primaries is discretely distributed in a range between 10$^{12}$ eV and 10$^{16}$ eV in steps of
0.5 in $\it\log E$. For the both layouts, the core position of any shower is uniformly distributed at all over the array surface.
The surface of the array is divided into 25 square pixels, each one with surface area of 49 $m^{2}$, and for any shower the core position is set in the
center of each pixel in order to increase the statistics of simulated EASs and also to cover different zones of the array.
Also a wide square grid including maximum number of 2809 square pixels is considered to study showers arriving inside and outside of the array surface while fired
the detectors. \\

Another set of showers includes 12000 proton and alpha showers, with the same previous composition, distributed over a continuous energy range from 2$\times$10$^{14}$ eV to 4$\times$10$^{14}$ eV according to a
power law with the spectral index of $\it \gamma$ = -2.7 have been generated to estimate the
angular resolution of the array. This energy range is selected according to the results of the first simulated showers set which reveal a significant
number of recorded events in this range in comparison with other bins of energy. In this set, the arrival direction of showers is selected from 0$^{\circ}$ to 60$^{\circ}$ for zenith angle in steps of 5$^{\circ}$ and from 0$^{\circ}$ to 360$^{\circ}$ for azimuth angle.\\

The response of the detectors to gammas as secondary particles is calculated by means of GEANT4 detector simulation toolkit~\cite{h}.
The energy deposition of gammas in the scintillation detectors is computed using simulated air showers generated by CORSIKA which are used as input for
detailed GEANT4 simulations. 

\section{Trigger probability function}
The trigger probability of a ground based array depends on several independent physical parameters: i) the characteristics of the primary cosmic ray which
initiates an air shower, e.g., energy and mass of the primary, ii) the type of the array detectors, iii) the trigger condition used to detect air showers, iv)
the array layout, v) the geometry of the incoming shower, e.g., its incidence zenith angle and distance of shower core to the array center~\cite{i}.
To show these dependencies, the trigger probability function, $\it P(r,E)$, has been presented as a fraction of showers in the energies of interest $\it(E)$ which fulfill the trigger condition in
different bins of shower core distance from the array center ($\it r$), as

\begin{equation}
\label{eq:1}
P(r,E)=\frac{N_{trigger}(r,E)}{N_{incident}(r,E)}.
\end{equation}

Where $N_{trigger}(r,E)$ and $N_{incident}(r,E)$ are number of showers with energy $\it E$ which fulfill the trigger condition at distance $\it r$ and total number of incident showers with given parameters $\it r$ and $\it E$, respectively. \\

The trigger probability function is an increasing function of energy, as the shower size gets bigger, and a decreasing function of distance from the shower core, as the density of particles gets smaller~\cite{j}. The trigger probability function changes between 0 and 1, which 0 happens for large shower core distances and 1 happens for the shower core distances close to the array center in high
energy showers. \\

In order to optimize the array size in a rectangular configuration, four different array sizes with detector spacings of 1.5 m, 3.5 m, 7 m and 14 m are studied.
It should be noted that the surface area of the site at the SUT campus lets the maximum detector spacing of 7 m. However, in the last phase site of the array will be extended. The trigger conditions are provided by randomly at least 4-fold (4 out of 20), 6-fold (6 out of 20), 10-fold
(10 out of 20) and finally fully triggering detectors. Due to the small surface of detectors, the fully triggering condition can provide more satisfactory statistical
sample of charged particles in the shower disc for future studies on the energy resolutions and mass composition. Although for the fully triggering condition,
the statistic of recorded events decreases significantly. The trigger conditions at least 4-fold and 6-fold achieve larger statistics of recorded events. Though, as discussed in the next section for the at least 10-fold condition, the angular resolution accuracy improves significantly in all zenith angles rather than at least
4-fold and 6-fold conditions. Thus, in the rectangular layout, we focus the analysis on the at least 10-fold condition. \\

\begin{figure}[tbp]
\centering 
\includegraphics[width=.49\textwidth]{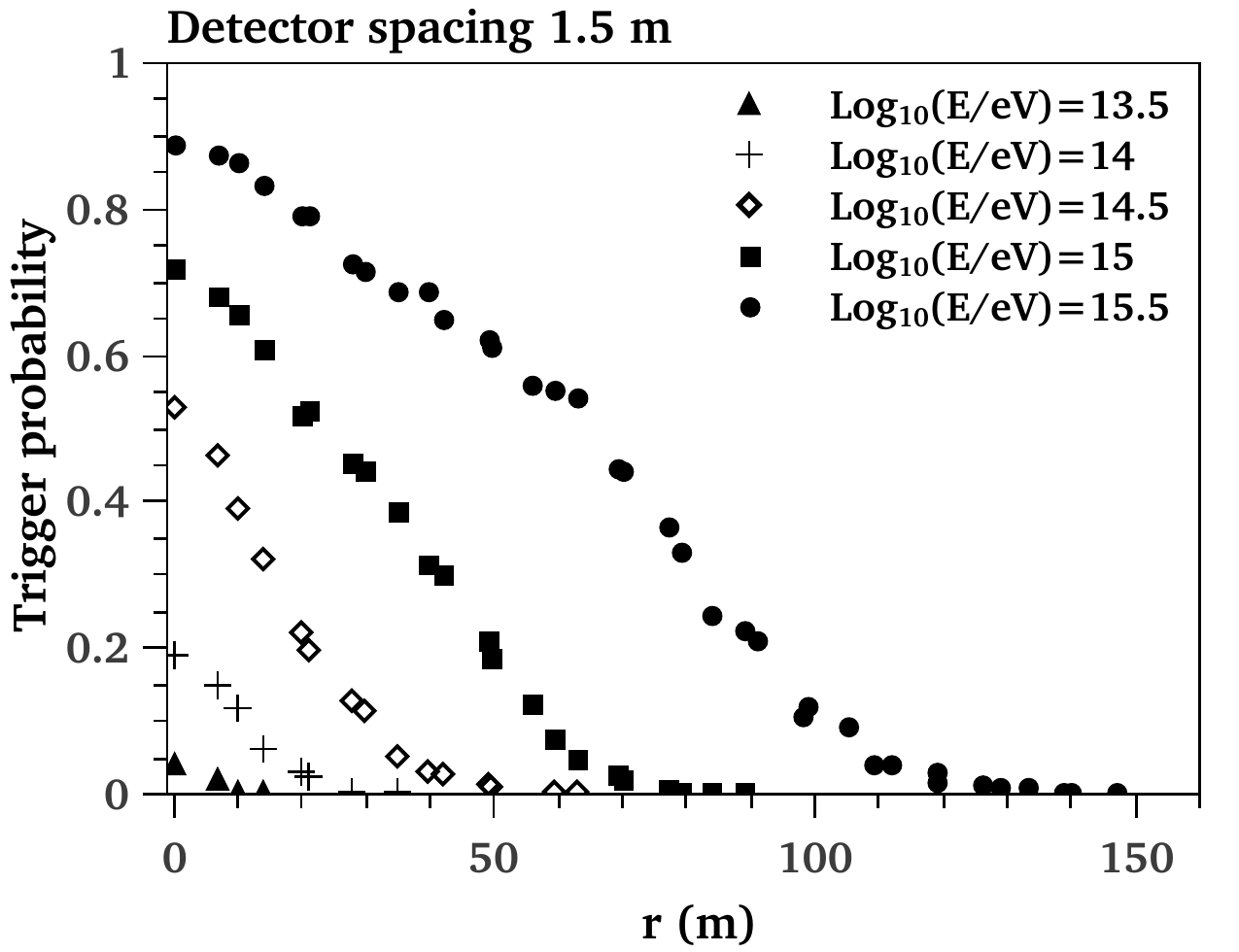}
\hfill
\includegraphics[width=.49\textwidth]{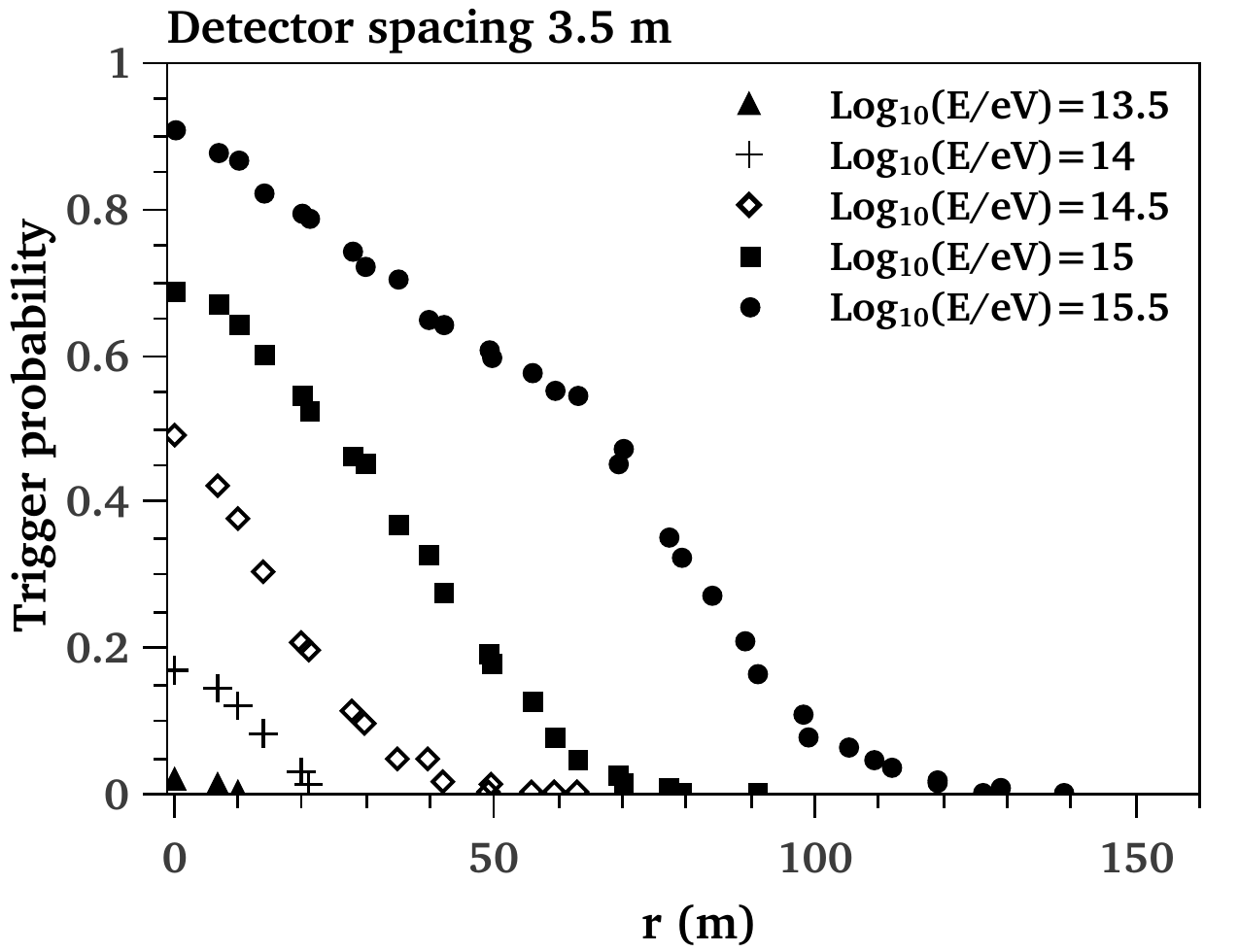}
\caption{\label{fig:2} Array trigger probability as a function of shower core distance to the array center $\it(r)$ for the 10-fold condition at different energies, from 3$\times$10$^{13}$ eV up to 3$\times$10$^{15}$ eV in steps of 0.5 in the logarithmic scale, for detector spacing of 1.5 m and 3.5 m in the rectangular grid.}
\end{figure}

\begin{figure}[tbp]
\centering 
\includegraphics[width=.49\textwidth]{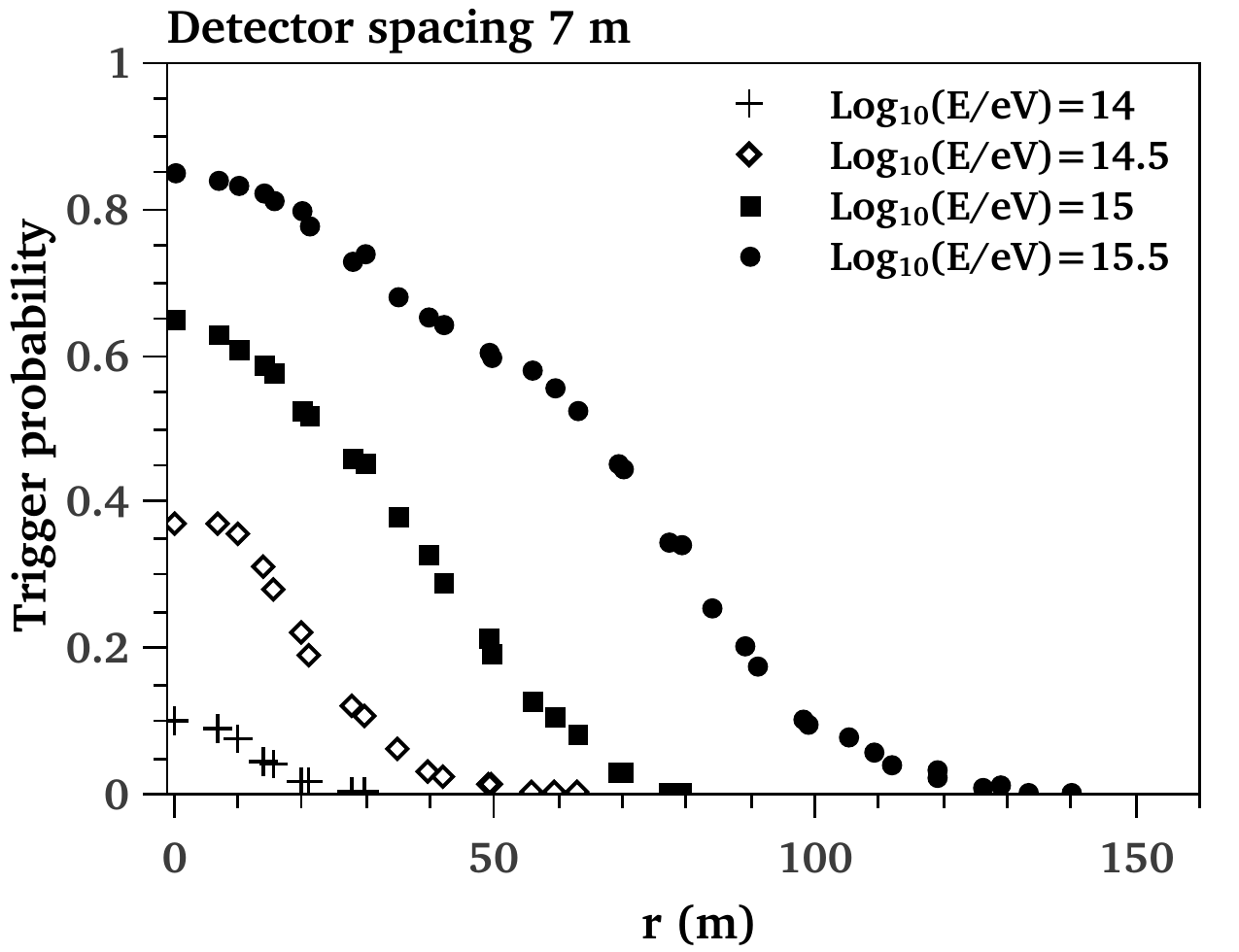}
\hfill
\includegraphics[width=.49\textwidth]{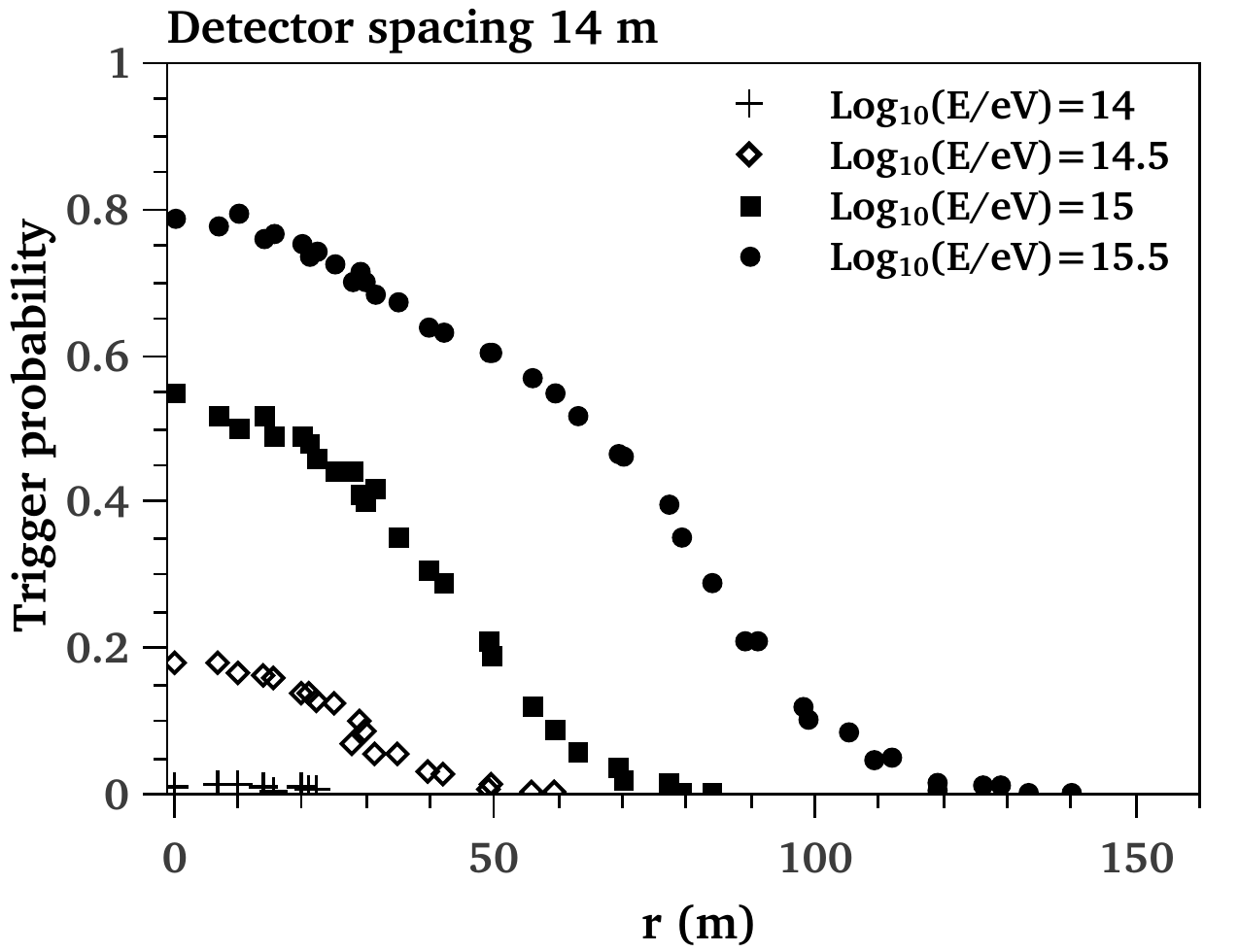}
\caption{\label{fig:3} Array trigger probability as a function of shower core distance to the array center $\it(r)$ for the 10-fold condition at different energies, from 10$^{14}$ eV up to 3$\times$10$^{15}$ eV in steps of 0.5 in the logarithmic scale, for detector spacing of 7 m and 14 m in the rectangular grid.}
\end{figure}

The trigger probability function from simulations for different given detector spacing in different bins of energy for the rectangular layout is shown in figures~\ref{fig:2} and~\ref{fig:3}. As can be seen, the maximum shower core distance ($\it r_{max}$) for detection in each energy is approximately independent of the detector spacing while it
increases with energy for a given detector spacing. \\  

Also the minimum energy of a shower that can be detected by the array obviously depends on the detector spacing so that for the smaller distances between neighboring detectors in the rectangular grid, efficiency on lower energies increases significantly. In other words, the
smaller array sizes allow to detect lower energies as the shower disc size becomes smaller. Therefore to focus on the $\it knee$, array size should be large
enough. \\

Beside the rectangular grid, a cluster layout is also studied. Through several implemented trigger conditions, those conditions with full triggered central cluster
and two surrounding triangles and or three surrounding triangles and or four surrounding triangles, (i.e., 5 out of 5 in the central cluster and 2 adjacent triangles, 5 out of 5 in the central cluster and 3 adjacent triangles,
and finally 5 out of 5 in the central cluster and 4 adjacent triangles) show higher efficiencies. In addition, their trigger probability are approximately the same. \\

\begin{figure}[tbp]
\centering 
\includegraphics[width=.49\textwidth]{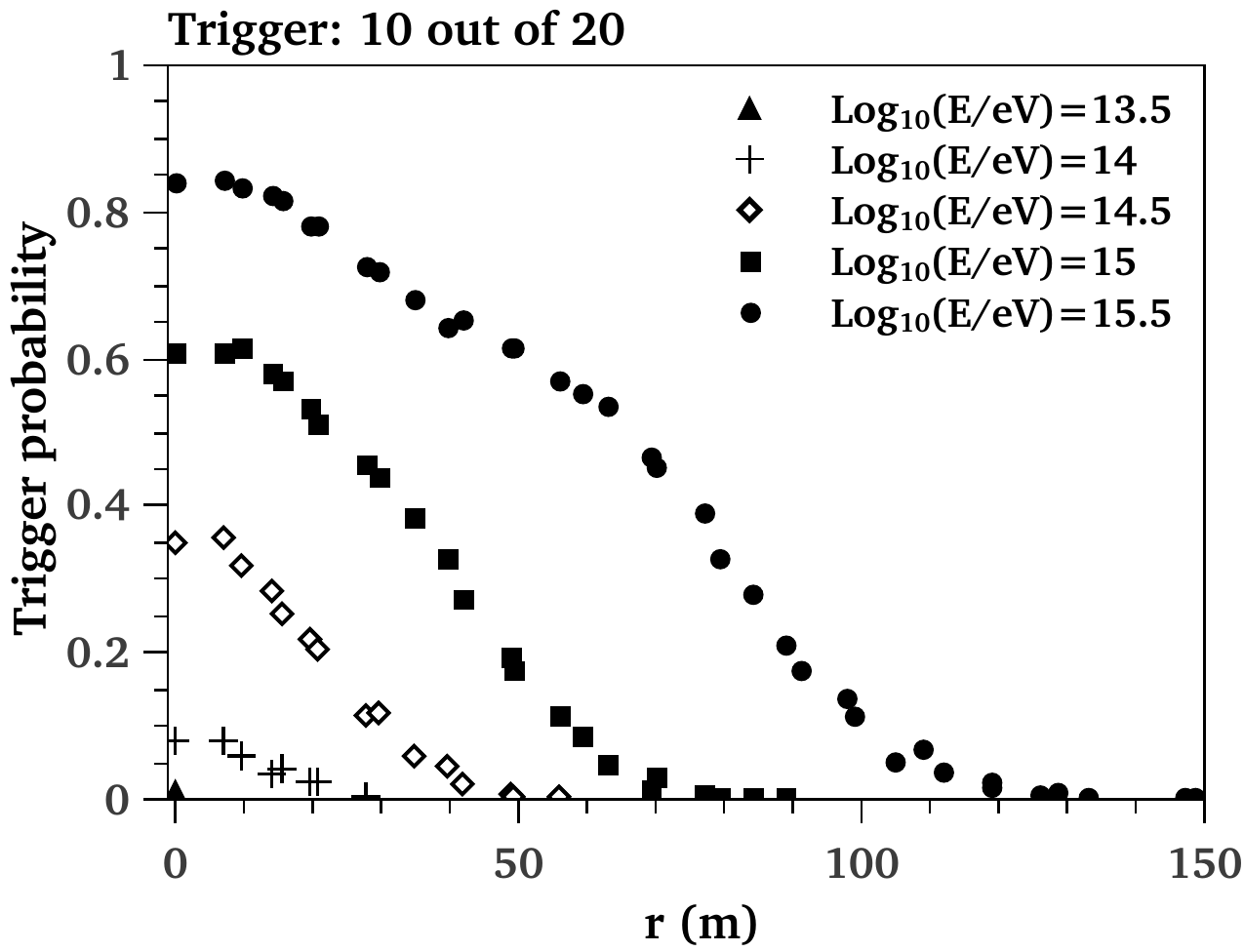}
\vfill
\includegraphics[width=.49\textwidth]{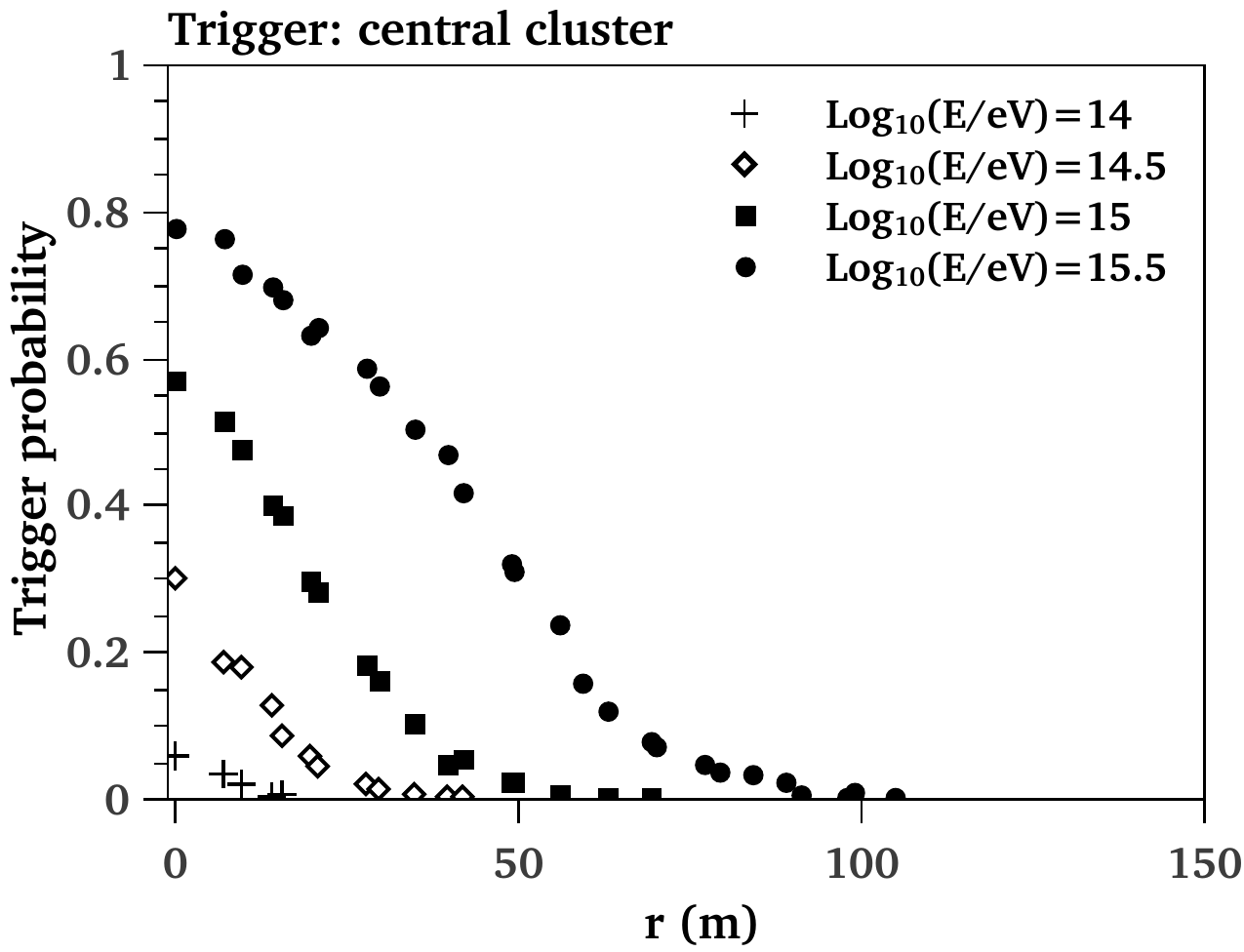}
\hfill
\includegraphics[width=.49\textwidth]{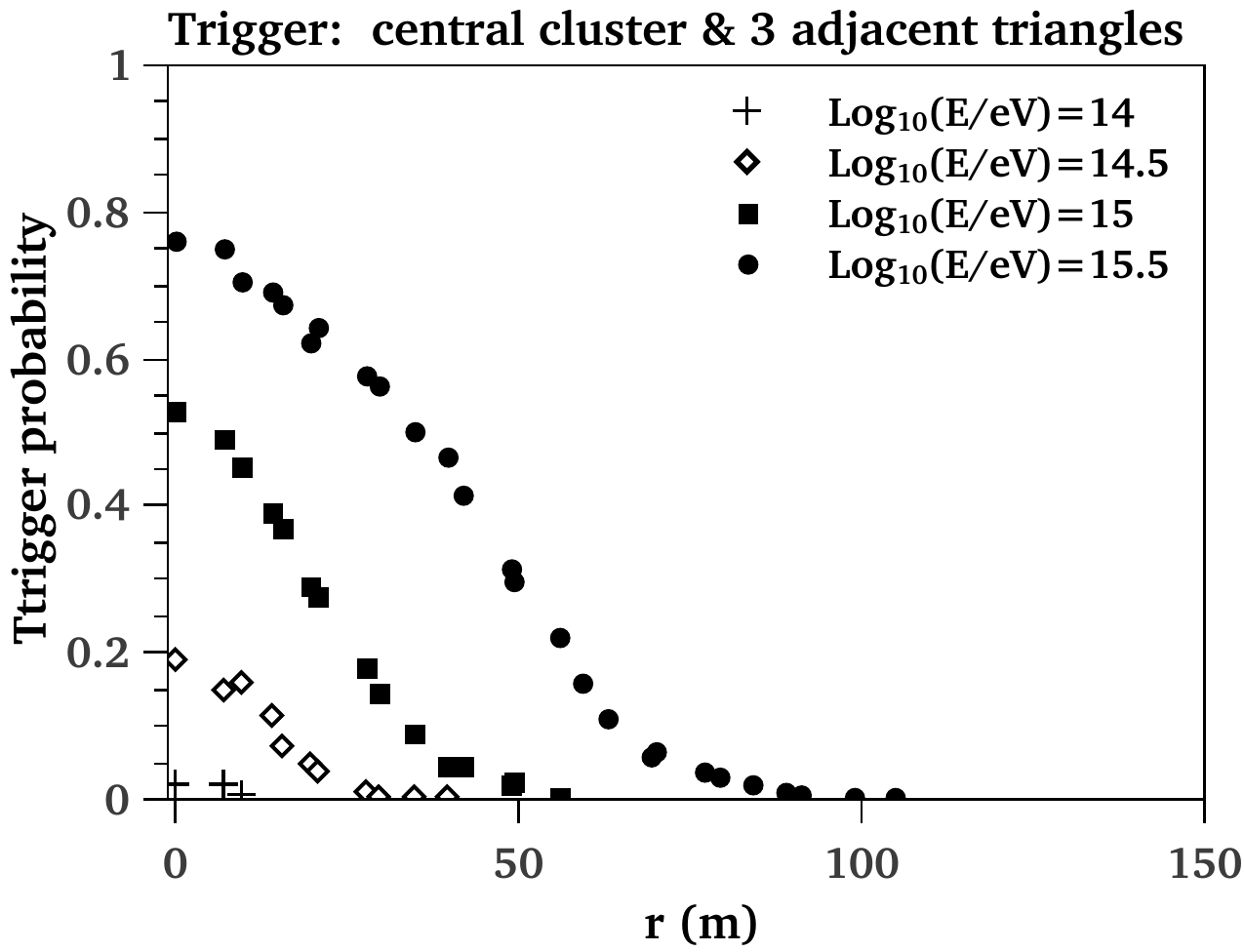}
\caption{\label{fig:4} Array trigger probability as a function of shower core distance to the array center ($\it r$) at different energies for triggering 10 out of 20 (top), 5 out of 5 in the central cluster (left-down), 5 out of 5 in the central cluster and 3 adjacent triangles (right-down).}
\end{figure}

Therefore, in triggering selection, these three trigger conditions are selected: randomly 10-fold, 5 out of 5 in the central cluster, and finally 5 out of 5 in the central cluster
and 3 adjacent triangles. As shown in figure~\ref{fig:4}, two last trigger conditions in the cluster layout decrease the maximum shower core distance ($\it r_{max}$) at different energies, e.g., $\it r_{max}$ is limited to 100 m for the primary energy of 3$\times$10$^{15}$ eV by triggering 5 out of 5 in the central cluster. Also
the condition of randomly at least 10-fold triggering shows almost the same trigger probability function in the cluster layout as well as the rectangular grid with spacing 7 m between neighboring detectors.\\

As shown in figures~\ref{fig:3} and~\ref{fig:4}, the trigger probability function becomes almost 0.8 at the energy of 3$\times$10$^{15}$ eV for shower core distances close to the array center for detector spacing
of 7 m and 14 m in the rectangular grid as well as the cluster layout. \\

In addition to the trigger probability function, number of triggered events per day in different bins of energy ($\it N(E,E+\triangle E)$) that fulfill the implemented trigger condition for different array layouts can be calculated by a convolution of the trigger probability function $\it P(r,E)$ and differential flux of primary cosmic rays $\it \Phi(E)$ (per area, time, solid angle and energy) (eq.\eqref{eq:2}),

\begin{equation}
\label{eq:2}
N(E,E+\triangle E)=\int_{S}\int_{E}^{E+\triangle E} P(r,E) \Phi(E)~dS~dE\int d\Omega \int dt
\end{equation}

where the intervals are selected according to the default initial conditions in the simulations. $\it S$ is the surface of a square grid which considered to study showers arriving inside and outside of the array while triggered it. $\it \Phi(E)$ (eqations \eqref{eq:3:1}, \eqref{eq:3:2} and \eqref{eq:3:3}) is a primary all-particle spectrum in the $\it knee$ region determined with the Tibet air shower array which is located at an almost ideal atmospheric depth for this energy range and is highly instrumented~\cite{j}. In the energy range below the $\it knee$, for $\it E$<5.62$\cdot10^{14}$~eV ,
\begin{subequations}\label{eq:3}
\begin{align}
\label{eq:3:1}
\Phi(E)=1.5\cdot10^{-20}(\frac{E}{5.62\cdot10^{14}})^{-2.60\pm0.04}~ [m^{-2}s^{-1}sr^{-1}eV^{-1}], 
\end{align}

and above the $\it knee$, for $E>7.08\cdot10^{15}~eV$,
\begin{equation}
\label{eq:3:2}
\Phi(E)=1.2\cdot10^{-23}(\frac{E}{7.08\cdot10^{15}})^{-3.00\pm0.05}~ [m^{-2}s^{-1}sr^{-1}eV^{-1}],
\end{equation}

and in the $\it knee$, at $E=1.78\cdot10^{15}~eV$,
\begin{equation}
\label{eq:3:3}
\Phi(E)=6.7\cdot10^{-22}~ [m^{-2}s^{-1}sr^{-1}eV^{-1}].
\end{equation} 
\end{subequations}

\begin{figure}[tbp]
\centering 
\includegraphics[width=.49\textwidth]{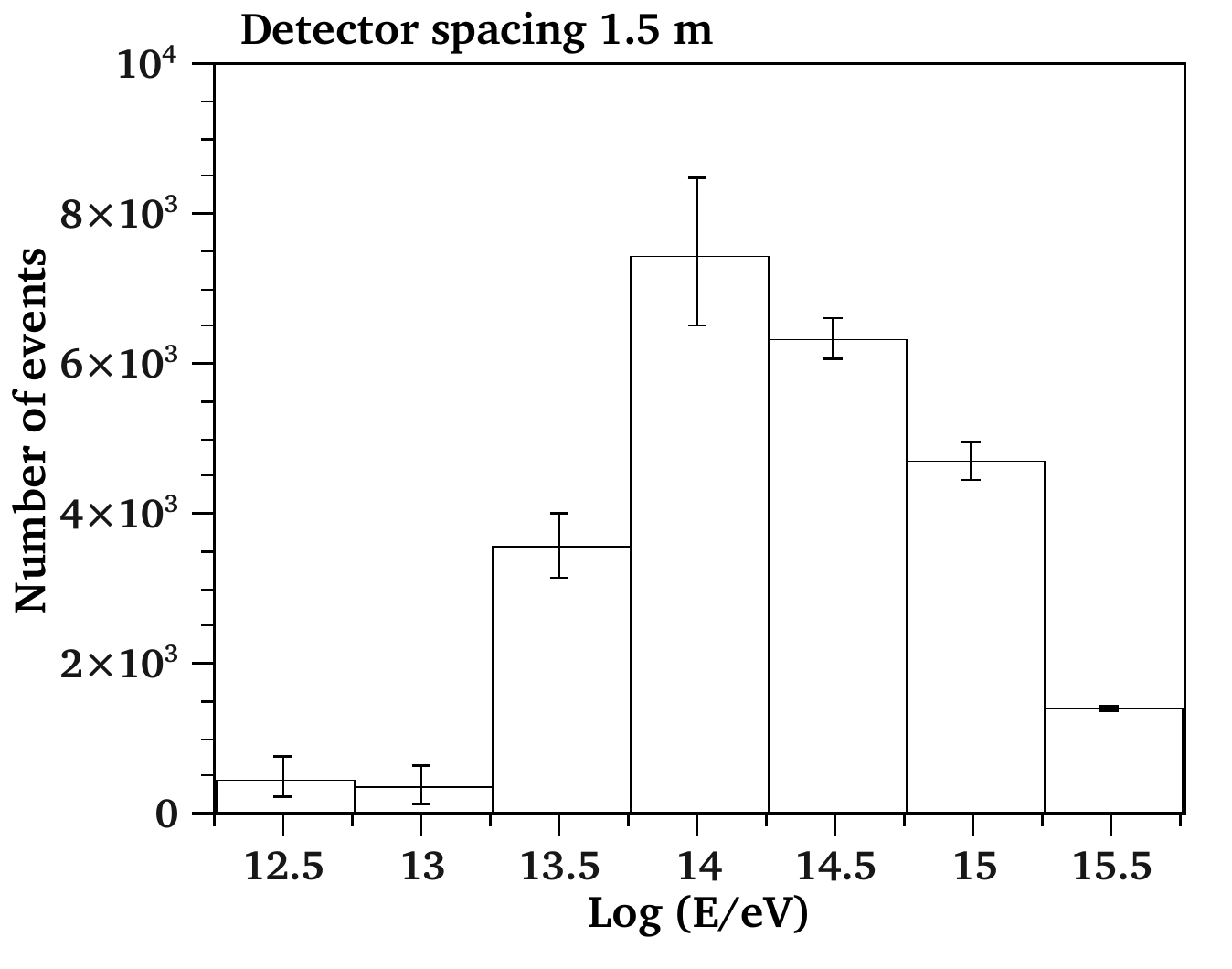}
\hfill
\includegraphics[width=.49\textwidth]{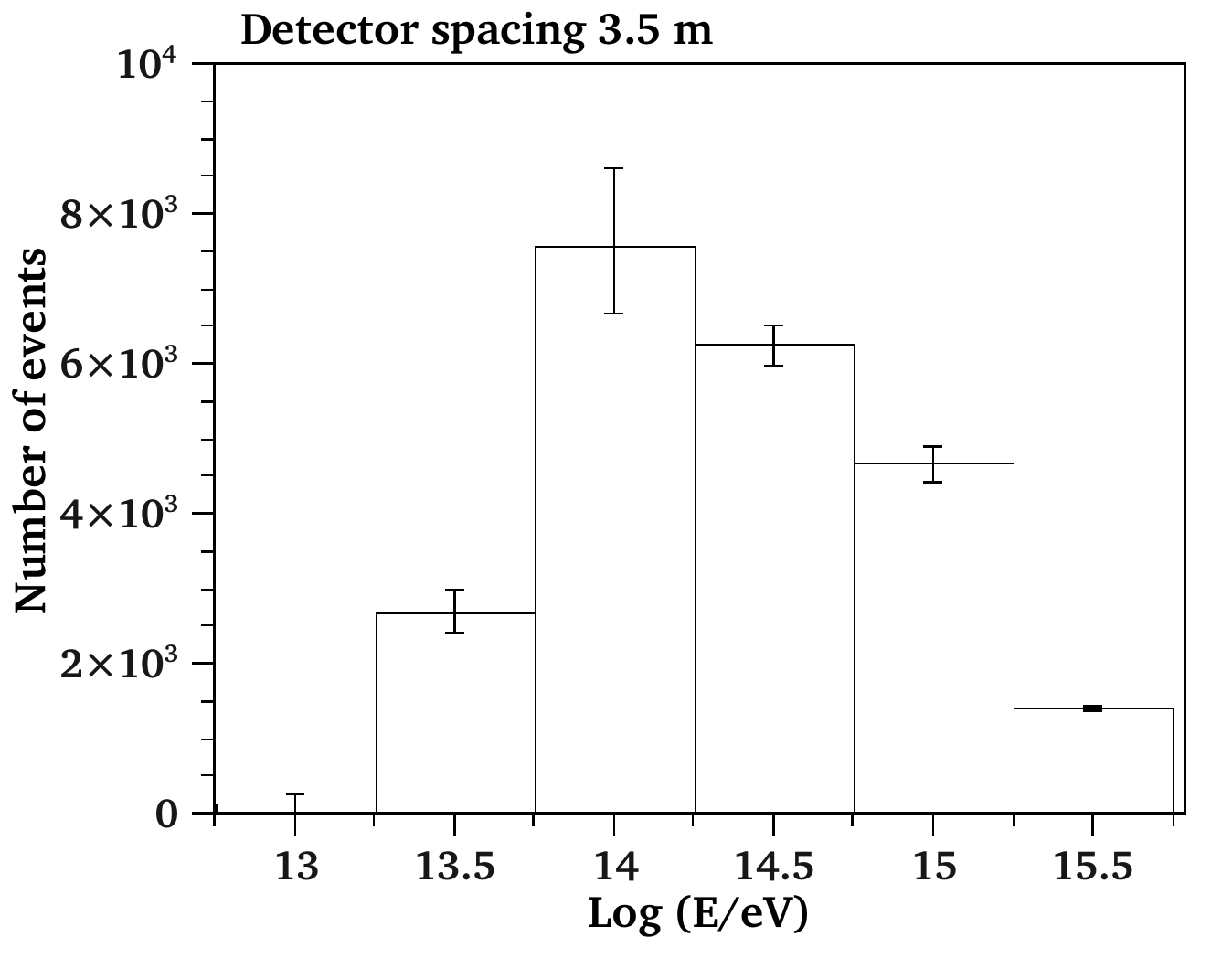}
\vfill
\includegraphics[width=.49\textwidth]{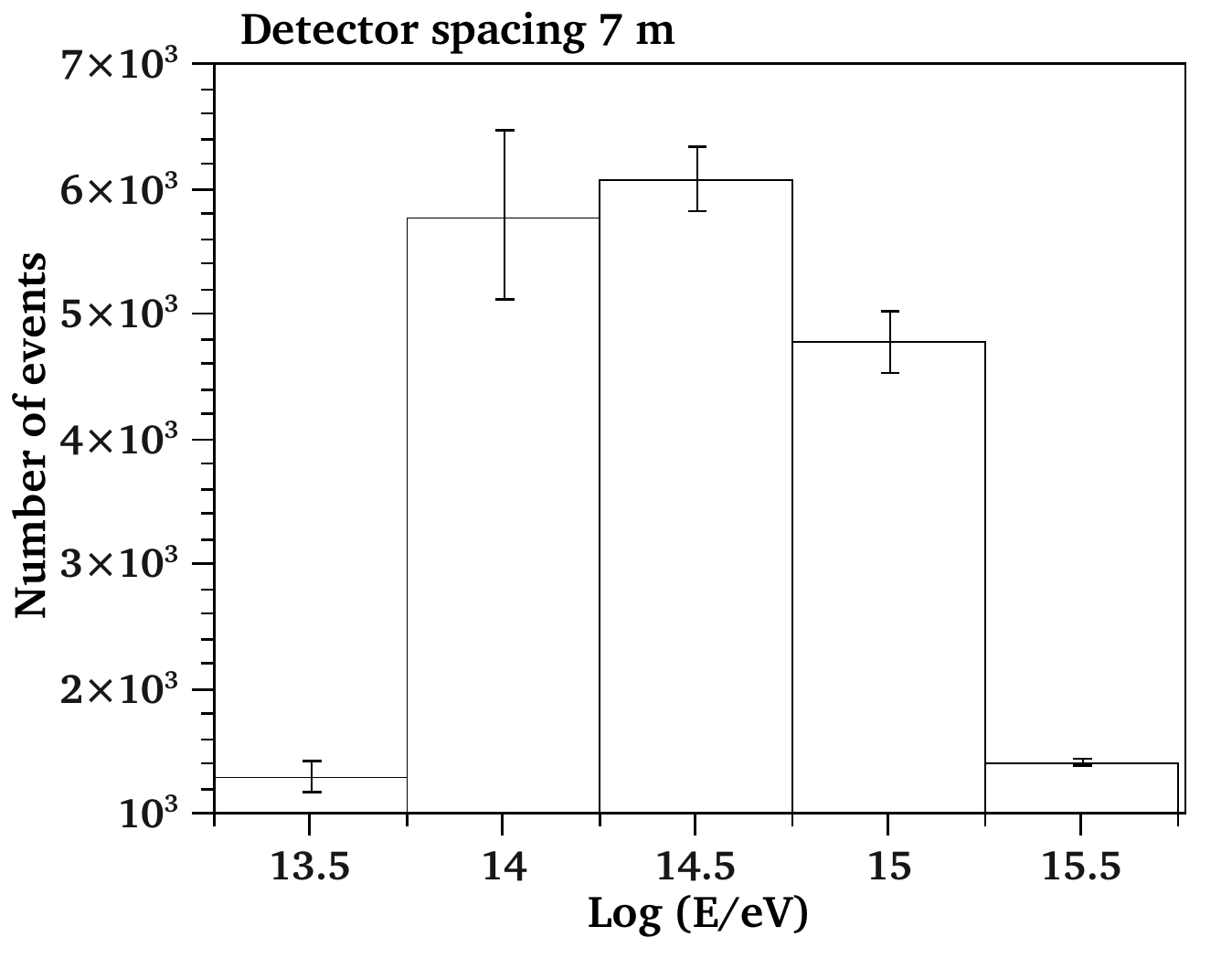}
\hfill
\includegraphics[width=.49\textwidth]{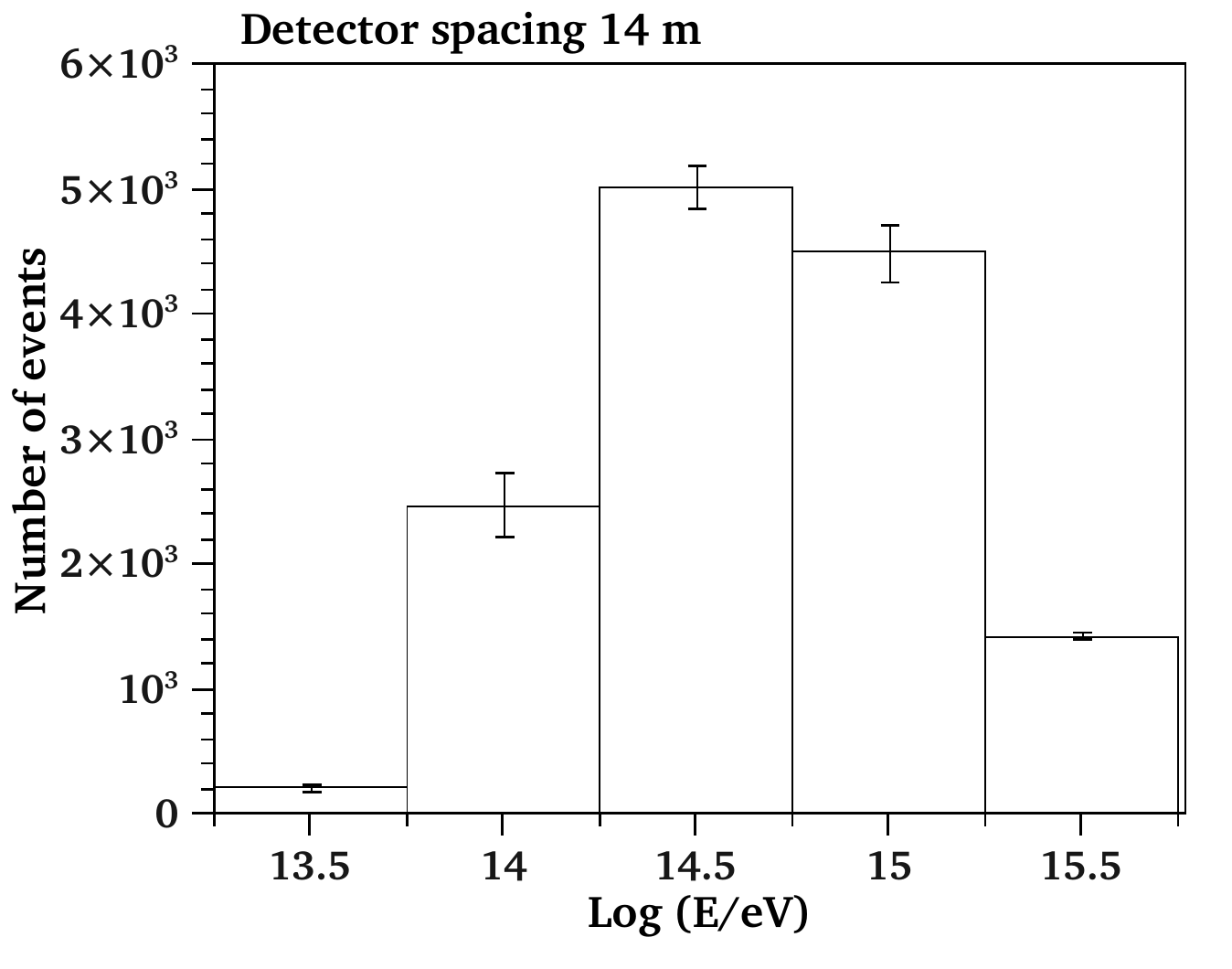}
\caption{\label{fig:5} Energy distribution of triggered events per day for different spacing between the neighboring detectors in the rectangular layout for the 10-fold condition.
}
\end{figure}

Figure~\ref{fig:5} shows the energy distribution of triggered events per day that fulfill the 10-fold condition for different array sizes in the rectangular layout. The results show an energy dependence of maximum triggered events on the detector spacing so that it changes from 10$^{14}$ eV in the small array sizes (detector spacing of 1.5 m and 3.5 m) to 3$\times$10$^{14}$ eV in the large enough array sizes (detector spacing of 7 m and 14 m).\\

\begin{figure}[tbp]
\centering 
\includegraphics[width=.49\textwidth]{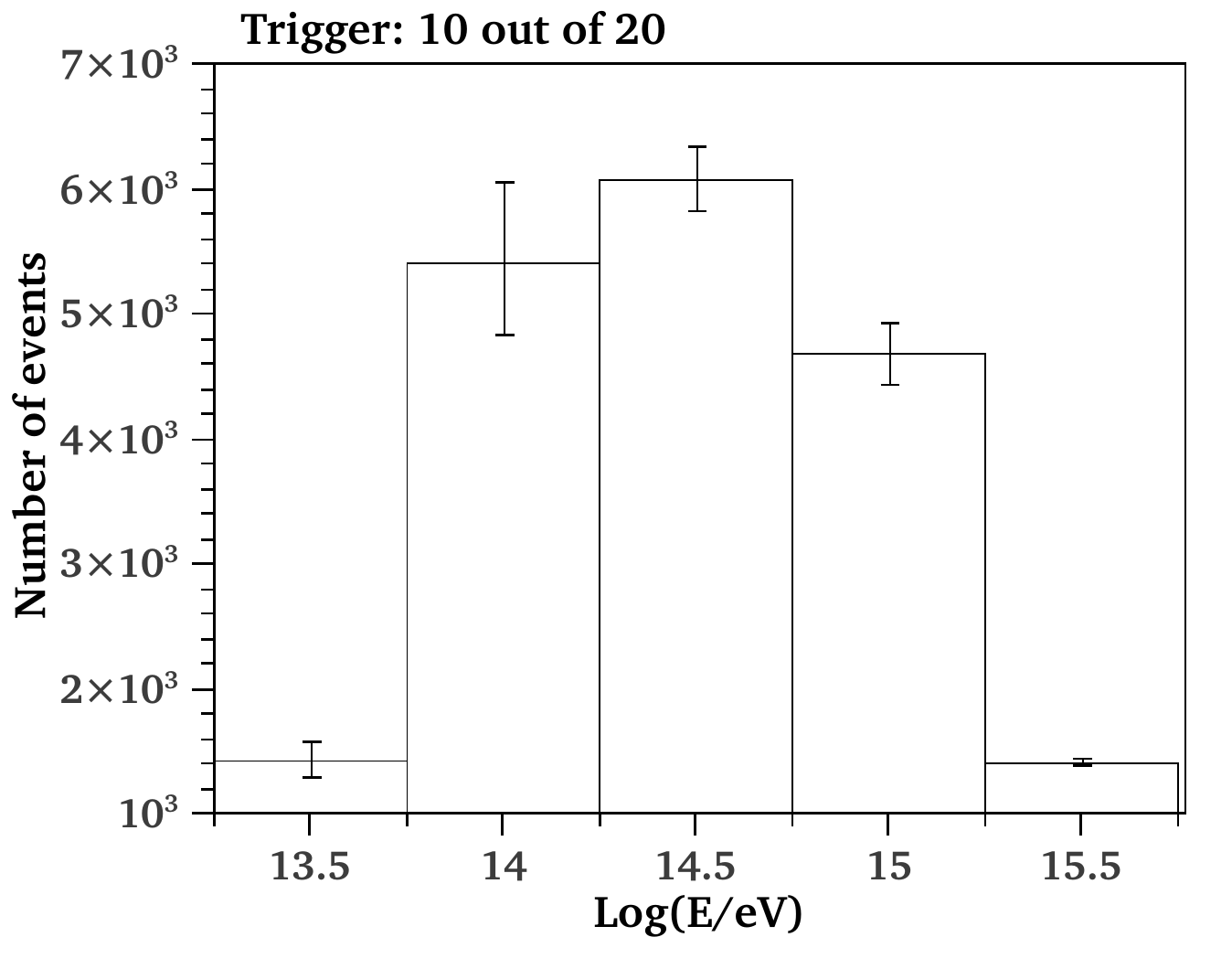}
\hfill
\includegraphics[width=.49\textwidth]{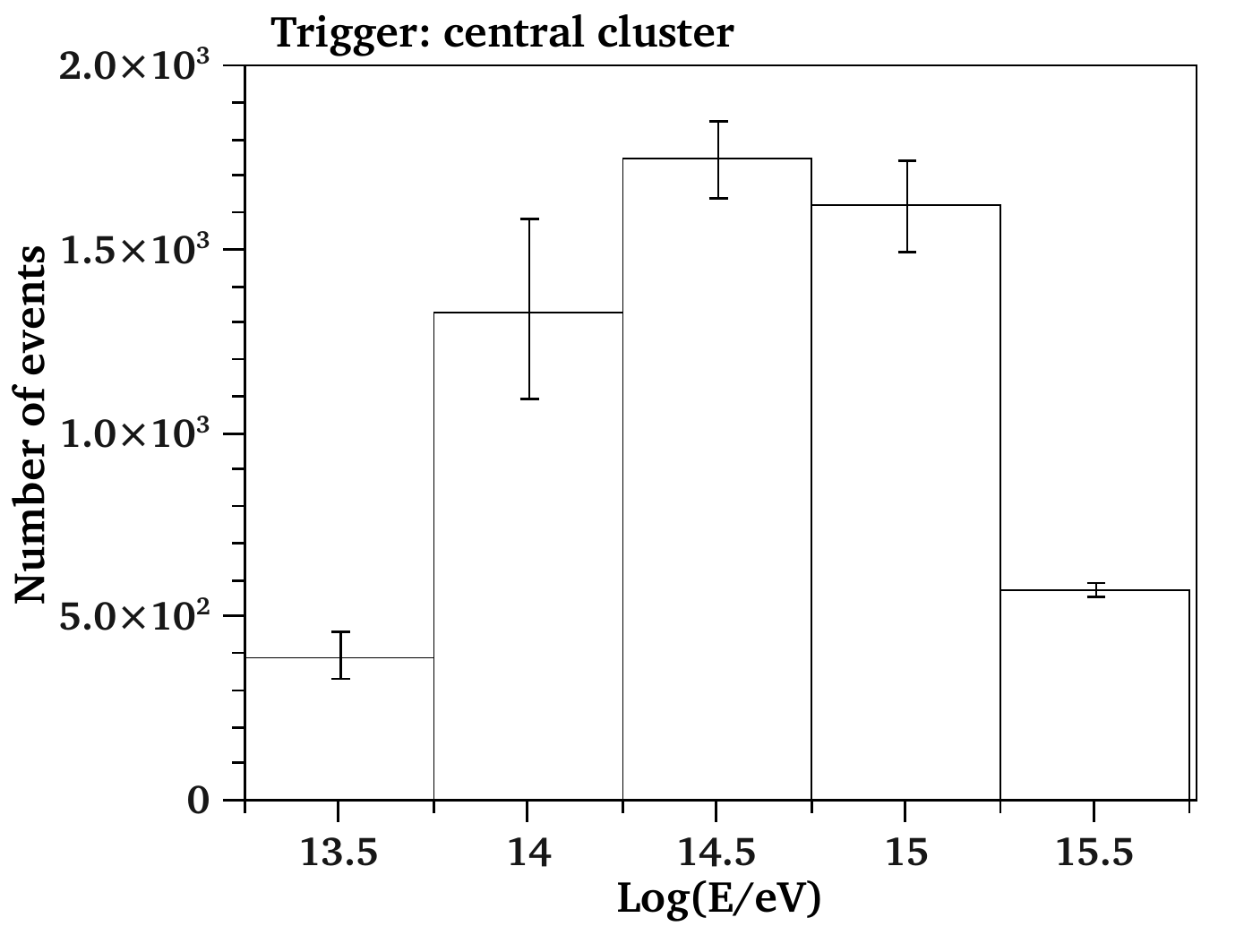}
\vfill
\includegraphics[width=.49\textwidth]{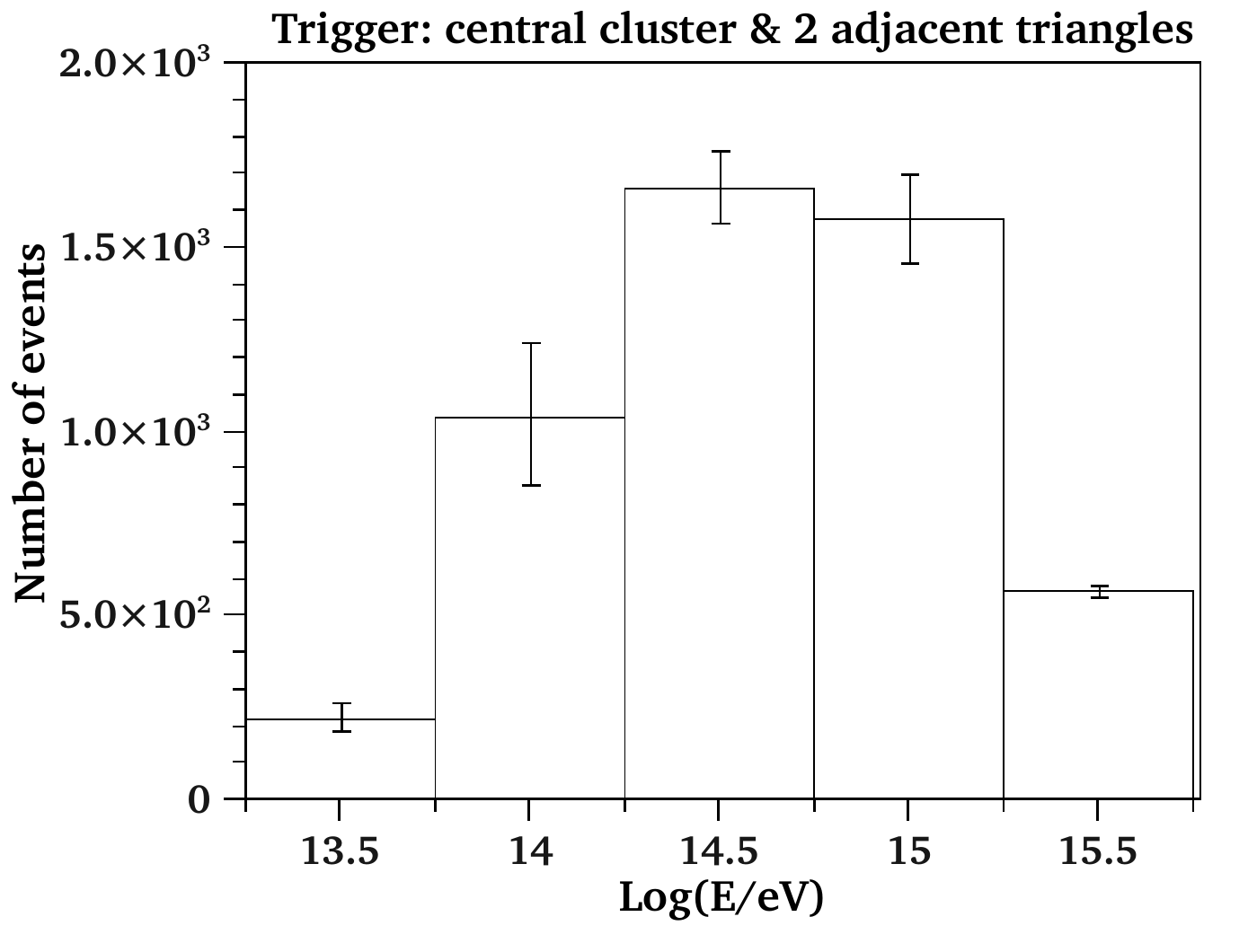}
\hfill
\includegraphics[width=.49\textwidth]{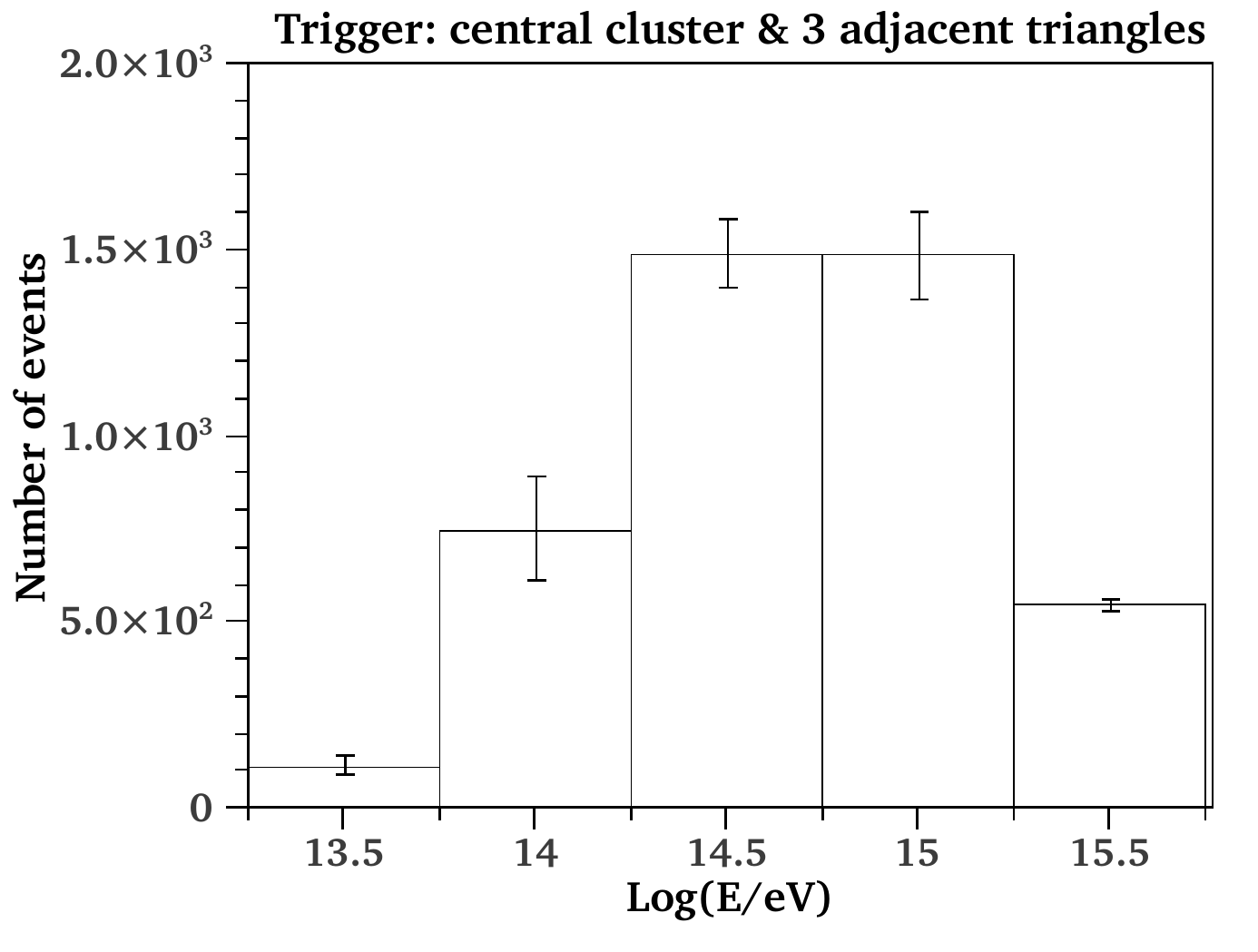}
\caption{\label{fig:6} Energy distribution of triggered events per day for different trigger conditions in the cluster layout.
}
\end{figure}

The energy distribution of triggered events per day for different trigger conditions in the cluster layout is shown in figure~\ref{fig:6}. As can be seen, the maximum happens in 3$\times$10$^{14}$ eV so that the array size is the same as the rectangular grid with detector spacing 7 m. It should be noted that the errors are due to systematic uncertainties of the cosmic ray spectrum (eqations \eqref{eq:3:1}, \eqref{eq:3:2} and \eqref{eq:3:3}) which is used in the Monte Carlo simulations to estimate the number of triggered events. \\

\section{Angular resolution}
The angular resolution of the Alborz-I is discussed by considering the second shower set including 12000 proton and alpha showers (with the same previous composition) distributed over a
continuous range from 2$\times$10$^{14}$ eV to 4$\times$10$^{14}$ eV with $\theta$ between 0$^{\circ}$ and 60$^{\circ}$ in steps of 5$^{\circ}$. This energy range is selected
according to receiving maximum number of events in $\it E$=3$\times$10$^{14}$ eV. \\

Using a chi-squared minimization algorithm the angular resolution of the array for different
trigger conditions and layouts is calculated. The angular resolution is directly related to the accuracy of timing measurement. Therefore to provide more realistic results in the simulation, the uncertainty on the timing
measurement is implemented and also core positions are randomly distributed at all over the array surface.\\ 

The timing measurement uncertainty by the Alborz-I instruments is calculated by:

\begin{equation}
\label{eq:3}
\sigma_{instr}=\sqrt{{\sigma_{det.}}^{2}+{\sigma_{elec.}}^{2}}\approx1.82~ns,
\end{equation}

where $\sigma_{elec.}=0.10~ns$ is the electronic uncertainty and $\sigma_{det.}=1.82~ns$ is the detector uncertainty~\cite{e}.\\

\begin{figure}[tbp]
\centering 
\includegraphics[width=.49\textwidth]{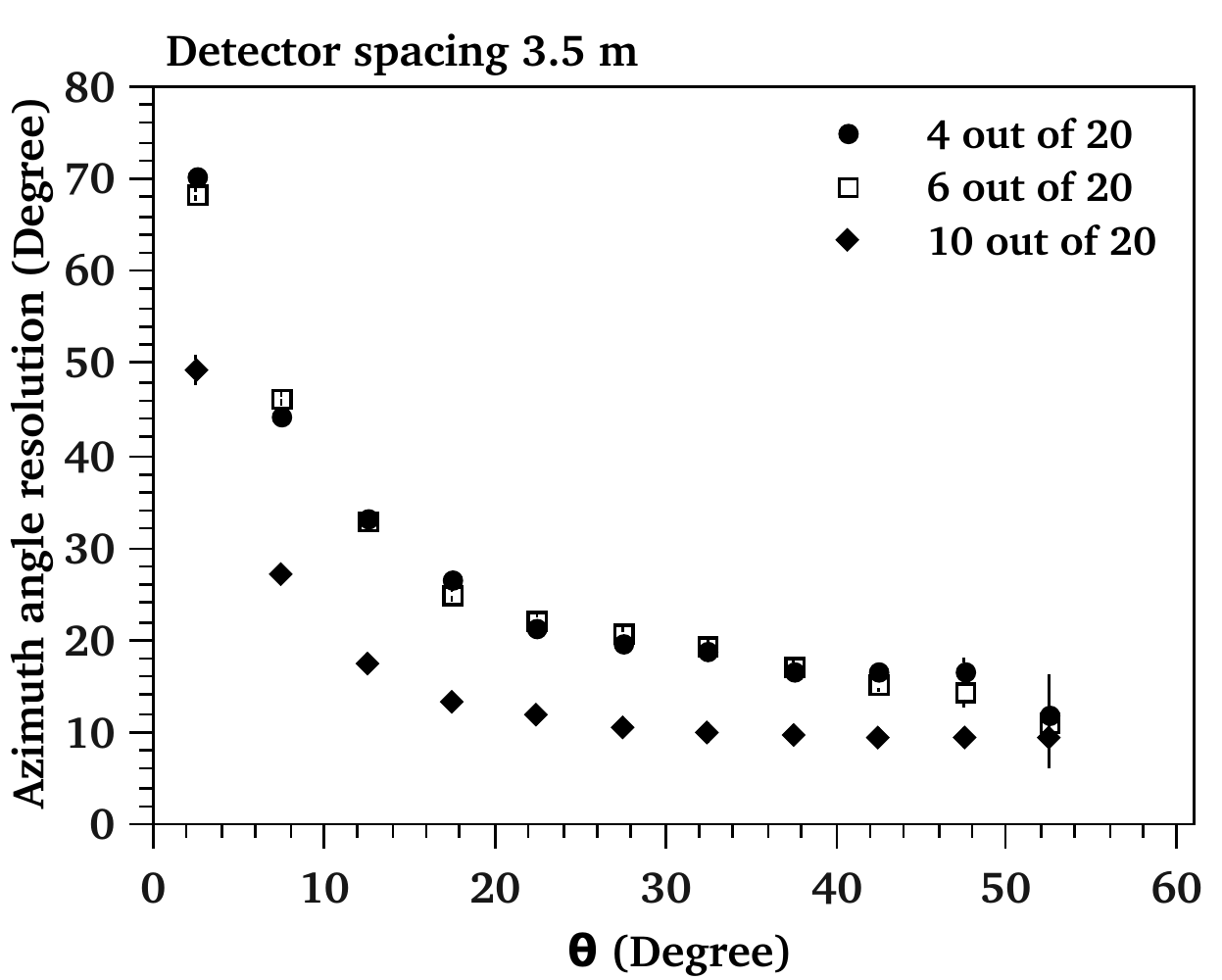}
\vfill
\includegraphics[width=.49\textwidth]{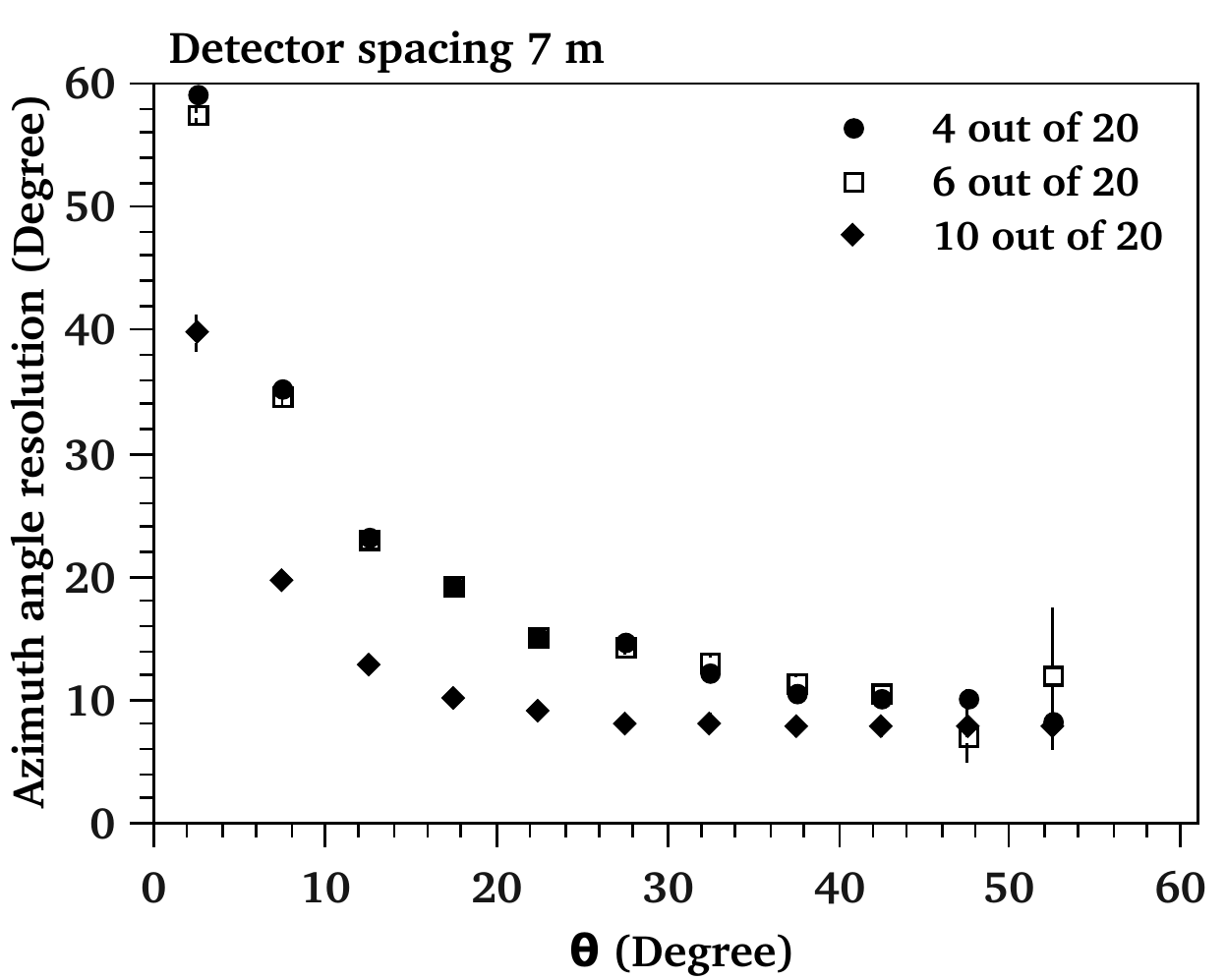}
\hfill
\includegraphics[width=.49\textwidth]{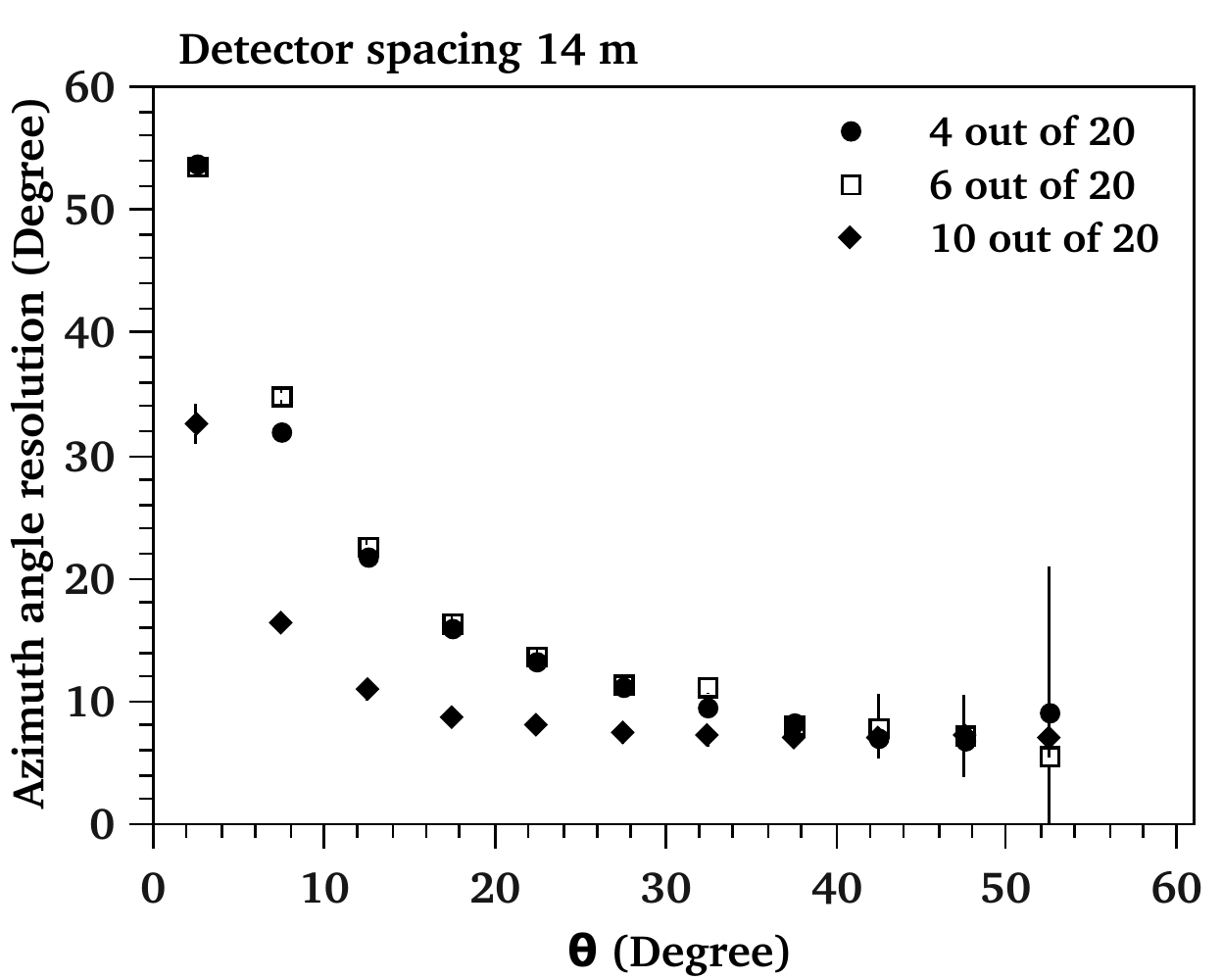}
\caption{\label{fig:7} Azimuth angle resolution for the detector spacing 3.5 m (top), 7 m (left-down) and 14 m (right-down) in the rectangular layout.
}
\end{figure}

\begin{figure}[tbp]
\centering 
\includegraphics[width=.49\textwidth]{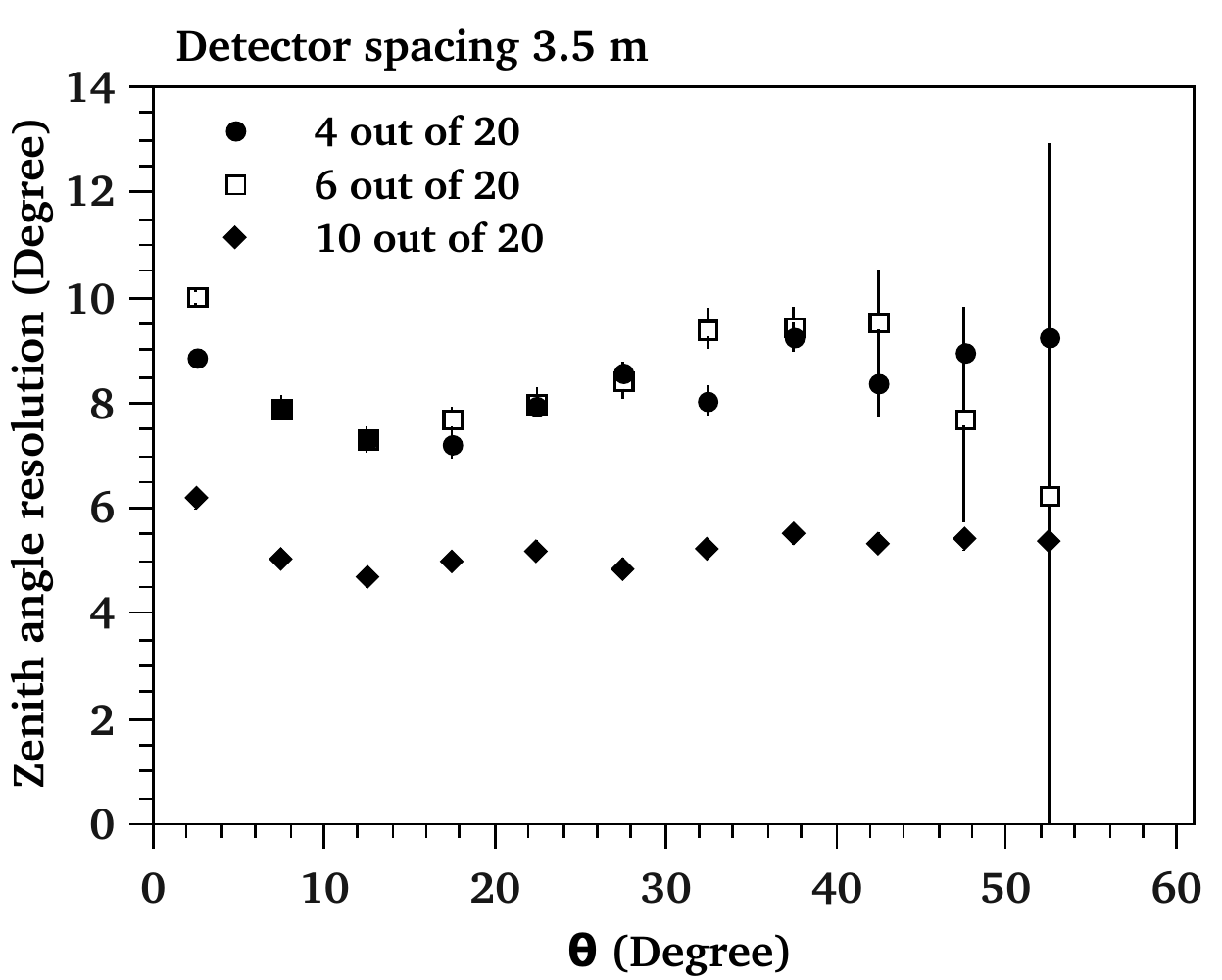}
\vfill
\includegraphics[width=.49\textwidth]{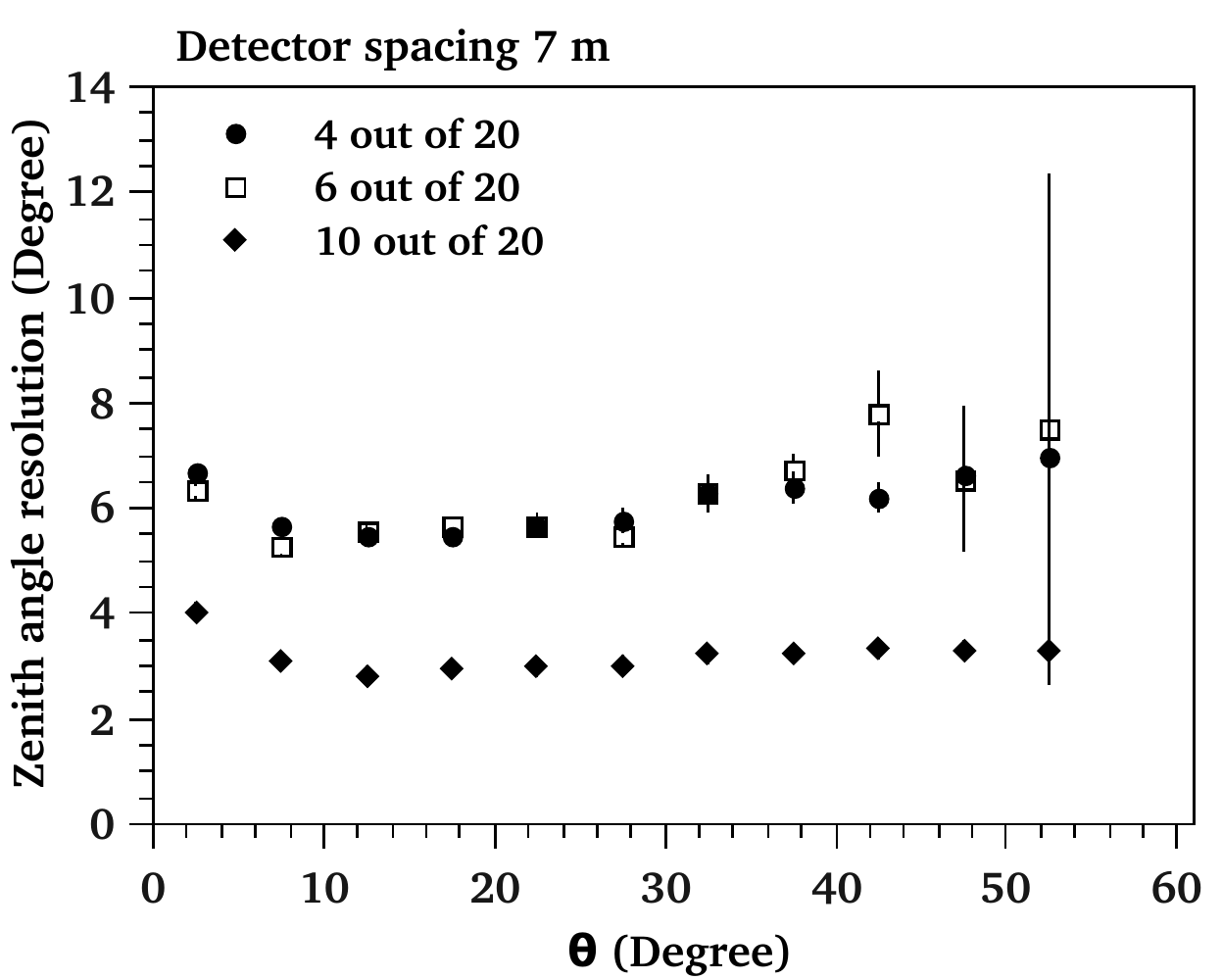}
\hfill
\includegraphics[width=.49\textwidth]{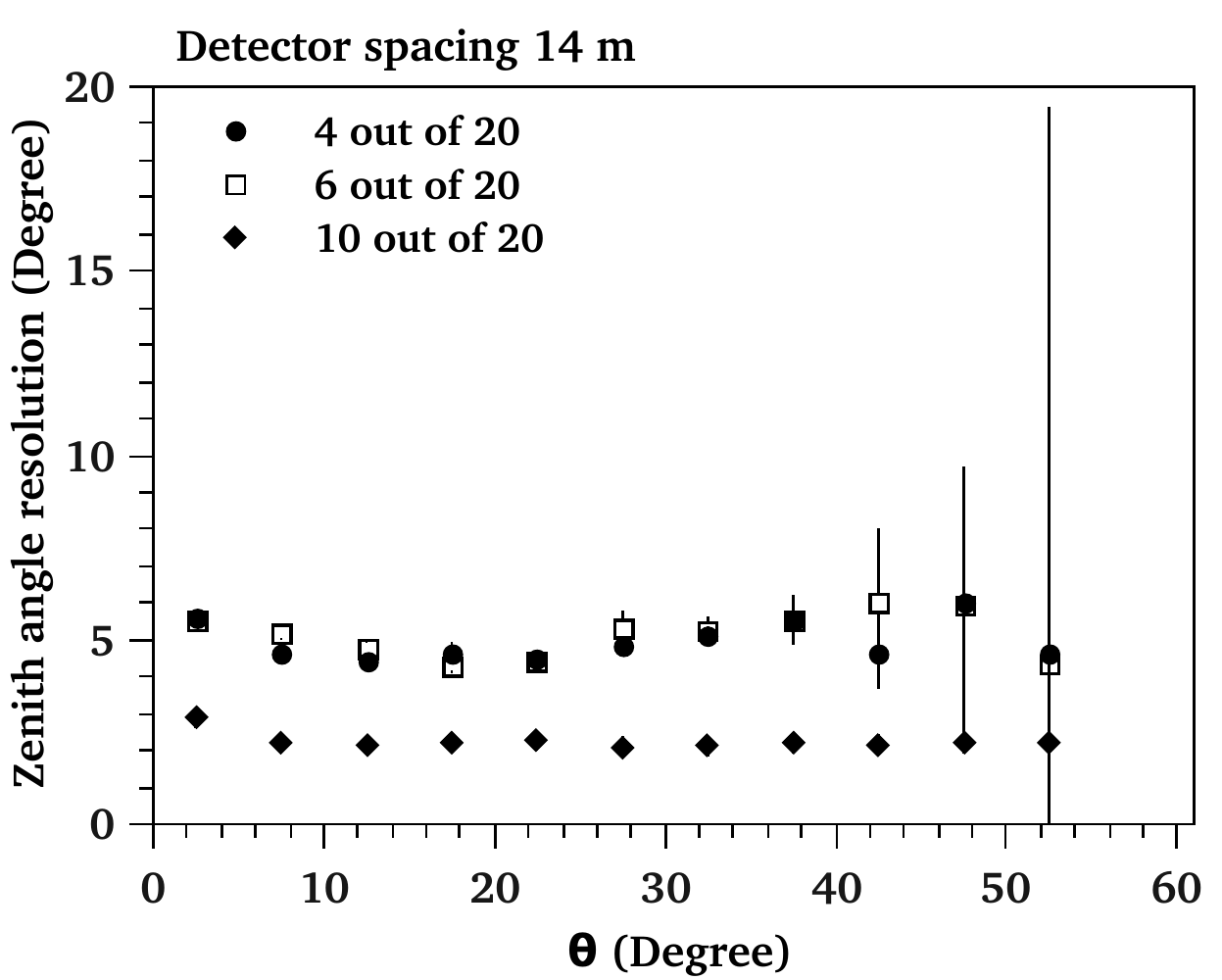}
\caption{\label{fig:8} Zenith angle resolution for detector spacing 3.5 m (top), 7 m (left-down) and 14 m (right-down) in the rectangular layout.}
\end{figure}

As shown in figures~\ref{fig:7} and~\ref{fig:8}, the angular resolution is obviously improved for triggering 10-fold in comparison with two other trigger conditions
for different array sizes in the rectangular grid. For the detector spacing 1.5 m due to small size of the array with respect to the size of simulated shower disc,
the angular resolution is minimized and not shown here.\\

\begin{figure}[tbp]
\centering 
\includegraphics[width=.49\textwidth]{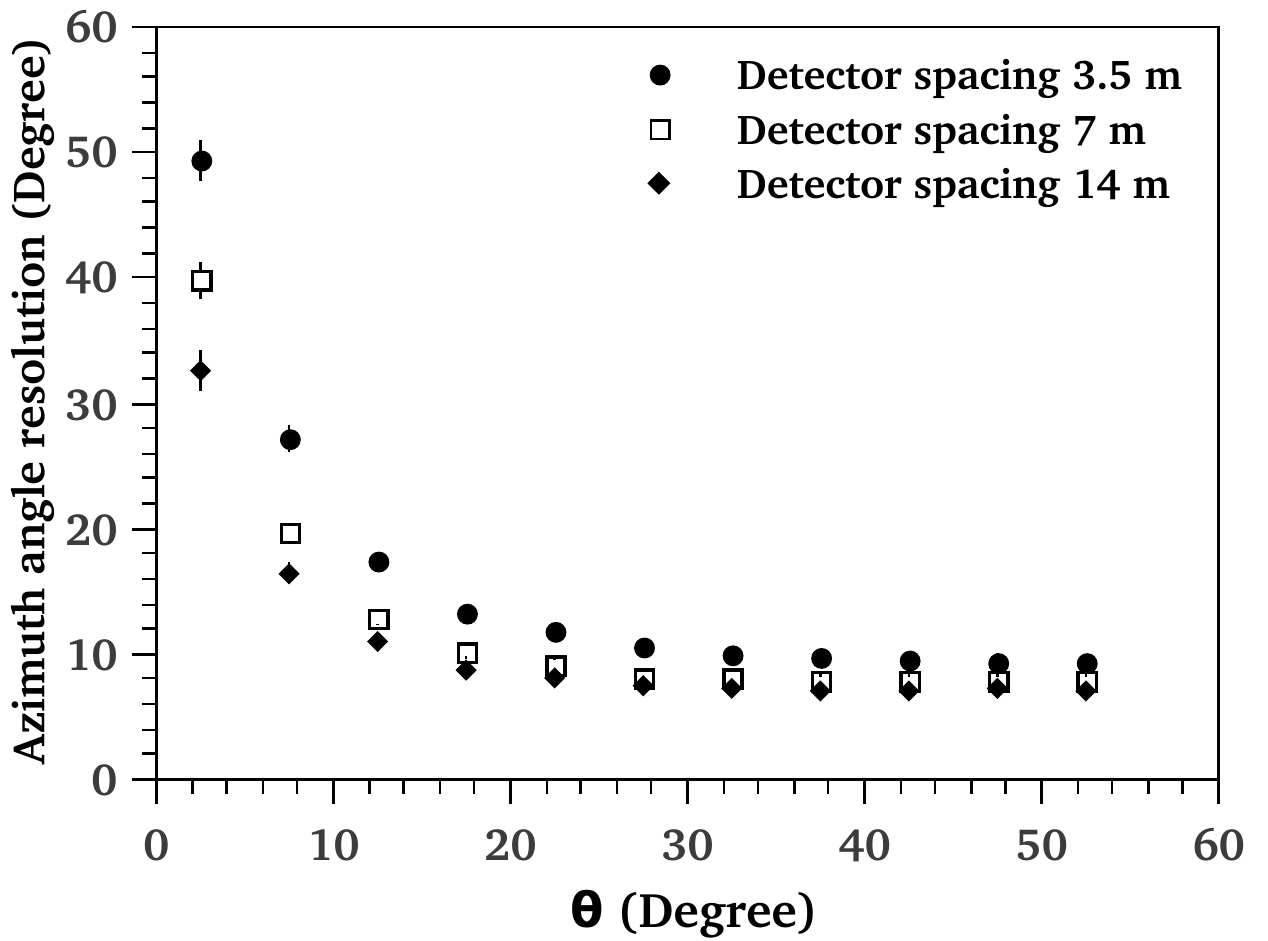}
\hfill
\includegraphics[width=.49\textwidth]{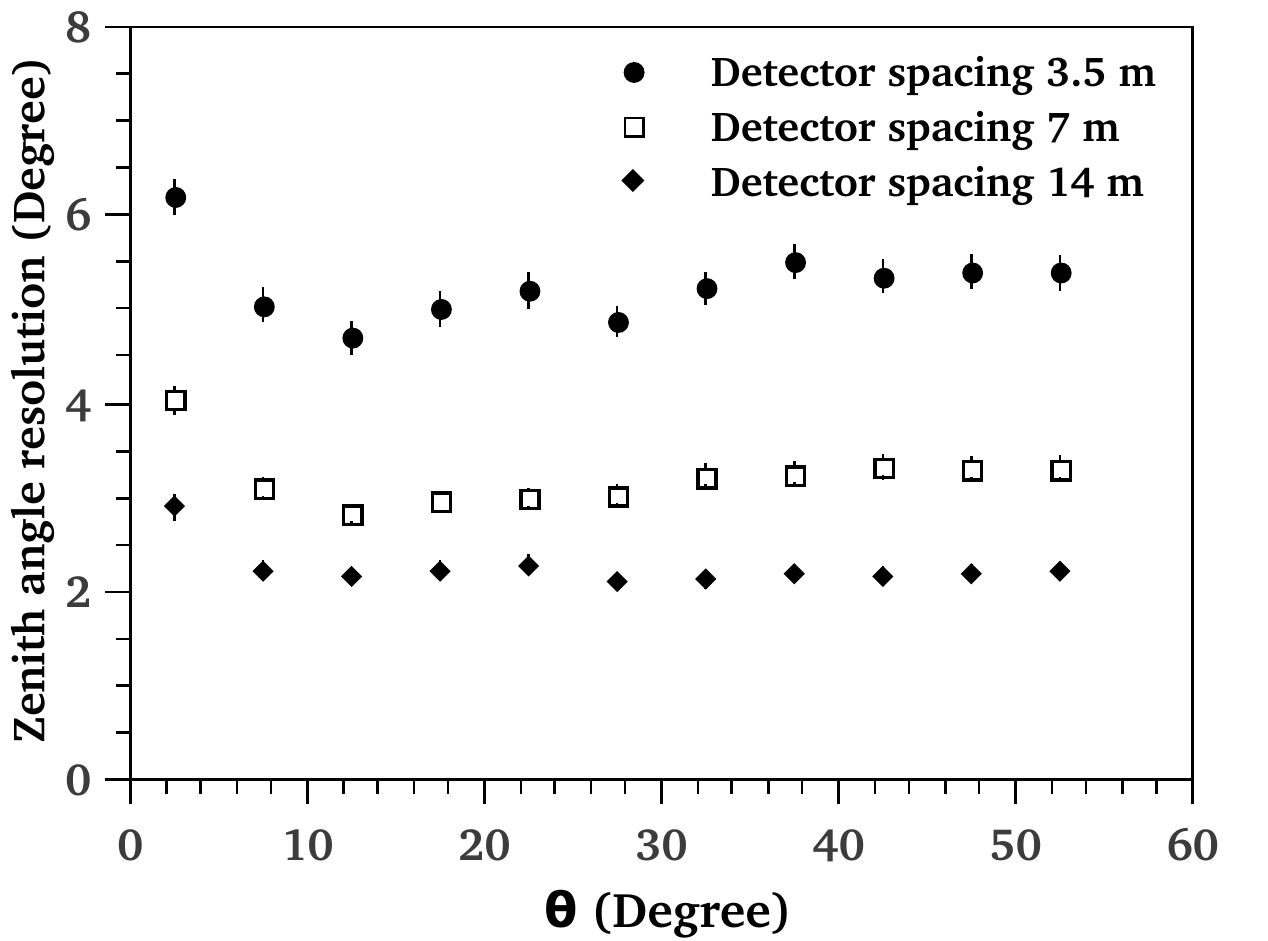}
\caption{\label{fig:9} Azimuth angle resolution (left) and zenith angle resolution (right) as a function of primary$^{,}$s zenith angle in the rectangular layout for the 10-fold condition.}
\end{figure}

Figure~\ref{fig:9} shows that the angular resolution is an increasing function of the array size and although the angular resolution improves for the detector
spacing 14 m, but there is a small difference between 7 m and 14 m detector spacing specially for zenith angles greater than 10$^{\circ}
$. In addition as mentioned above, the site of array limits detector spacing to 7 m.\\

\begin{figure}[tbp]
\centering 
\includegraphics[width=.49\textwidth]{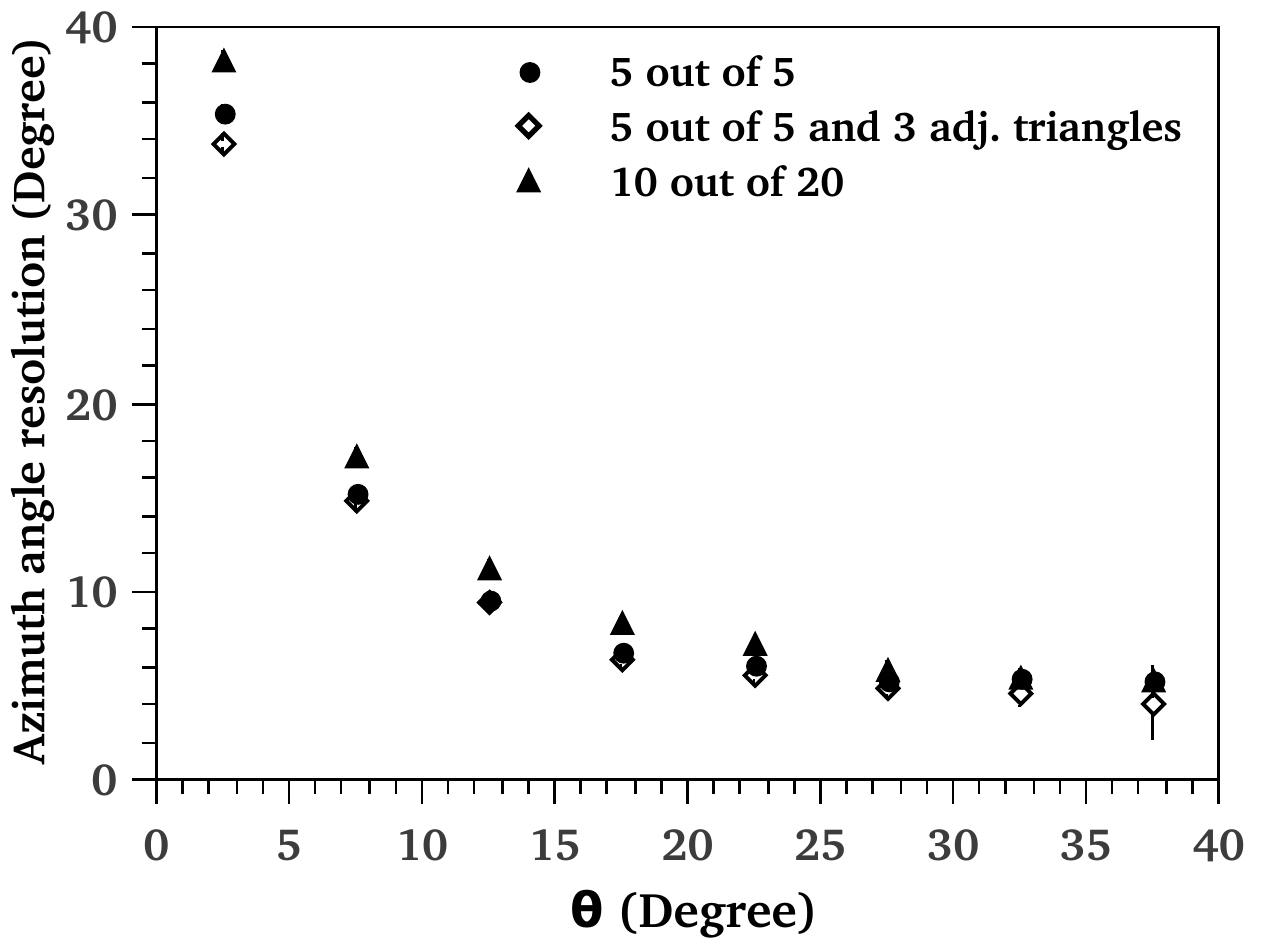}
\hfill
\includegraphics[width=.49\textwidth]{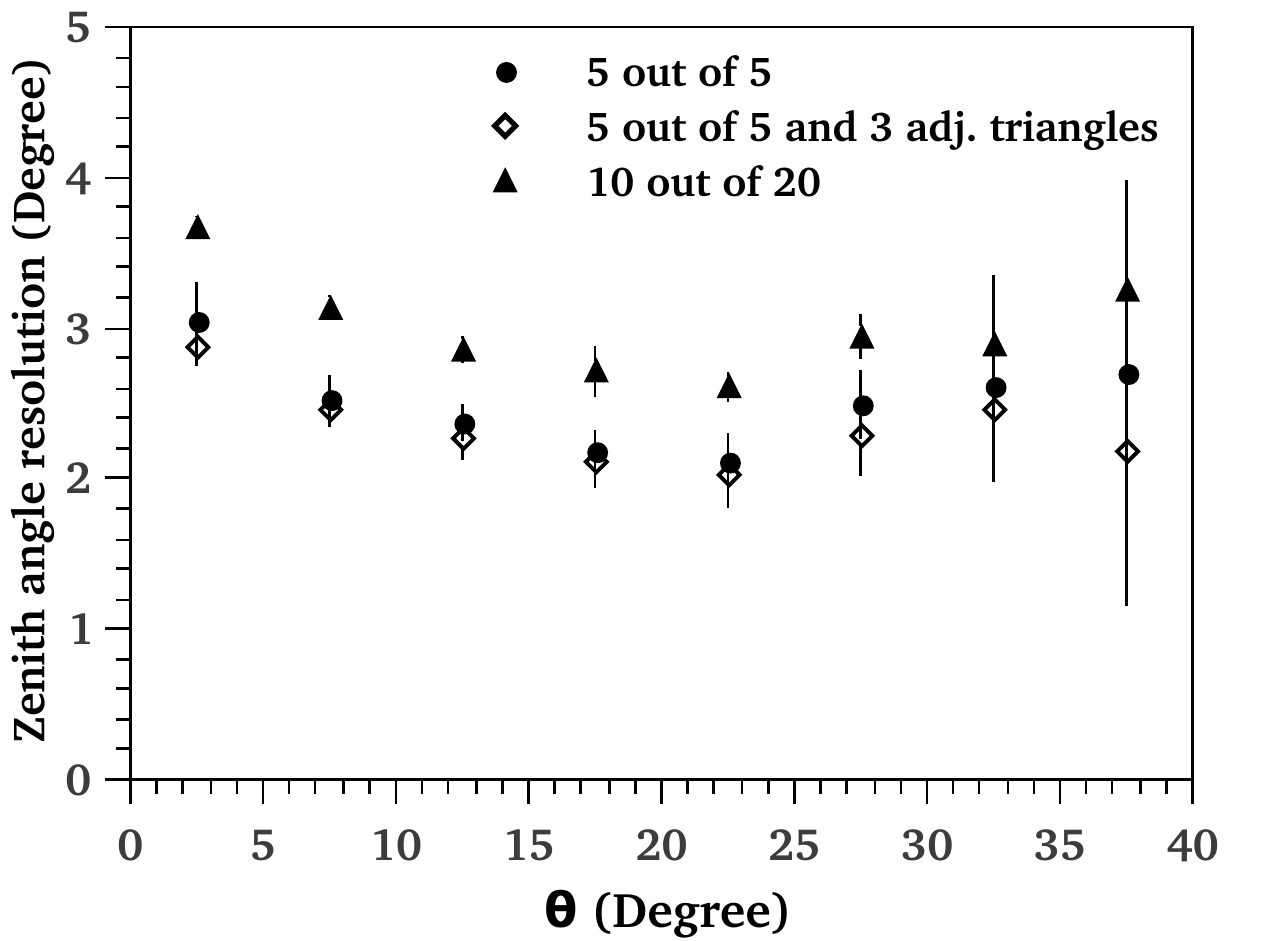}
\caption{\label{fig:10} Azimuth angle resolution (left) and zenith angle resolution (right) as a function of primary$^{,}$s zenith angle for the pentagon layout.}
\end{figure}

As illustrated in figure~\ref{fig:10}, the angular resolution improves in the pentagon layout in comparison with the rectangular grid with triggering 10-fold and
detector spacing 7 m. Through the three selected trigger conditions in the pentagon layout, those which include fully triggered central cluster show almost the
same angular resolutions. The errors represent statistical uncertainties as well as systematic uncertainties which is due to the time resolution of the array
instruments.\\

\section{Conclusion}
Monte Carlo studies based on the CORSIKA program on the Alborz-I array characteristics have been performed. The trigger probability
function, angular resolution and energy distribution of recorded events per day have been estimated for two layouts and different trigger conditions. \\

The array response, trigger conditions and shower development are summarized in the trigger probability function. The results indicate that although at lower energies the
trigger probability decreases significantly but due to high flux of cosmic rays, large number of events are recorded. Moreover at higher energies (around the $\it knee$ of
the cosmic ray spectrum), the array is more efficient.\\

In the rectangular layout, four different trigger conditions and four array sizes were considered. Study on the angular resolution showed that it obviously improves
as the array size and number of triggered detectors increase. But limitation in the surface area of the array and, the significant decrease in recording events for
fully triggering of the array (20-fold triggering condition) made us to consider the detector spacing 7 m and the triggering of at least 10-fold as a default condition to
compare with the results of the pentagon layout.\\

In the pentagon layout the implemented trigger conditions, which include fully triggered central cluster together with some of the adjacent triangles yield
approximately the same efficiency in different energy ranges. Energy distribution of recorded events for different trigger conditions in the pentagon layout
maximize at around 3$\times$10$^{14}$ eV. The angular resolution of the pentagon layout improves significantly in comparison with the rectangular grid with 7 m
detector spacing and 10-fold triggering conditon. 

\acknowledgments
We are very grateful to the Sharif High Performance Computing Center (HPCC) for all their services and technical supports. We thanks the members of the 
Cosmic Ray Laboratory, specially Mehdi Abbasian Motlagh and Masoume Rezaei for their assistance.

\end{document}